\documentclass[apj, twocolappendix]{emulateapj}
\usepackage{natbib}
\usepackage{sidecap}
\usepackage{verbatim}
\usepackage{mathtools}
\usepackage[caption=false]{subfig}
\usepackage{bm}
\usepackage{hyperref}
\usepackage{graphicx, amsmath, amsthm, amssymb,color}
\usepackage{CJK}
\usepackage{ulem} 

\def\boldxi{\mbox{\boldmath $\xi$}}
\def\boldv{\mbox{\boldmath $v$}}
\def\boldr{\mbox{\boldmath $r$}}
\def\boldnabla{\mbox{\boldmath $\nabla$}}

\def\boldF{\mbox{\boldmath $F$}}
\def\boldn{\mbox{\boldmath $n$}}
\def\boldp{\mbox{\boldmath $p$}}
\def\bnabla{\mbox{\boldmath $\nabla$}}

\def\Bruntfreq{Brunt-V{\"a}is{\"a}l{\"a}\,\,\,}
\def\refnew#1{(\ref{#1})}
\def\be{\begin{equation}}
\def\ee{\end{equation}}

\def\erg{\, \rm erg}
\def\erg{\, \rm erg}

\def\s{\, \rm s}
\def\km{\, \rm km}
\def\cm{\, \rm cm}
\def\g{\, \rm g}
\def\m{\, \rm m}

\def\icarus{Icarus}

\shorttitle{Oscillations of a giant planet}
\shortauthors{Wu \& Lithwick}

\begin{document}

\title{Memoirs of a giant planet}

\author{Yanqin Wu$^{1}$ and  Yoram Lithwick$^{2}$}
\affil{$^1$ Department of Astronomy \& Astrophysics, University of Toronto}
\affil{$^2$ CIERA, Department of Physics \& Astronomy, Northwestern University}

\begin{abstract}  
  Saturn is ringing, weakly.  Exquisite data from the {\it Cassini} mission
  reveal the presence of f-mode oscillations as they excite density
  waves in Saturn's rings. These oscillations have displacement
  amplitudes of order a metre on Saturn's surface.  We propose that
  they result from large impacts in the past.  Experiencing
  little dissipation inside Saturn on account of its weak luminosity,
  f-modes may live virtually forever; but the very ring waves that
  reveal their existence also remove energy from them, in $10^4$ to
  $10^7$ yrs for the observed f-modes (spherical degree $2-10$). We
  find that the largest impacts that arrive during these times
  excite the modes to their current levels, with the exception of the
  few lowest degree modes. To explain the latter, either a
  fortuitously large impact in the recent past, or a new source of
  stochastic excitation, is needed.
%
  We extend this scenario to Jupiter which has no substantial rings.
  With an exceedingly long memory of past bombardments, Jovian f-modes
  and p-modes can acquire much higher amplitudes, possibly explaining
  past reports of radial-velocity detections, and detectable by the
  {\it Juno} spacecraft.
\end{abstract} 

\section{Introduction}
\label{sec:intro}

Seismology has long been a valuable probe of the interiors of the
Earth and Sun.  But for giant planets, oscillations have not been
detected until recently.  \cite{stevenson1982saturn} first proposed
that the rings of Saturn can act like a giant seismograph. An
oscillating mode within Saturn produces density perturbations. These
produce an oscillating gravitational field which can launch a density
wave in Saturn's rings, at the location of the mode's Lindblad
resonance.  \cite{Marley91} and \cite{MP93} refined Stevenson's
proposal, showing that Saturn's prograde, sectoral (i.e., with
spherical harmonic integers $l=m$) f-modes will produce detectable features
in Saturn's C-rings, provided surface
displacements are over about 1 metre. These, \citet{MP93} argued, could explain some of
the features seen in the {\it Voyager} data that are un-associated
with any known satellites.
They also showed that f-modes with $l-m=1$ can produce
vertical bending waves in the rings, and those with larger values of
$l-m$ can also perturb the rings, albeit with smaller amplitude.


Such a scenario has been unambiguously confirmed recently.  By careful
analysis of the {\it Cassini} stellar occultation data,
\citet{HN13,HN14,HN18, French19} identified density waves associated
with all of the prograde, sectoral f-modes with $\ell \leq 10$.  The
association of an observed density wave with an internal mode is made
by observing both the number of spiral arms and the pattern frequency
of the wave, and then comparing with theoretical f-mode frequencies.
Frequencies calculated by \citet{Mankovich} show excellent agreement
with those inferred from the observed pattern frequencies \citep[also
see previous works by][]{Vorontsov78,Vorontsov81}.
In addition to density waves launched by sectoral f-modes, a number of
other ring waves have also been observed, including density waves due
to f-modes with $l-m=2$ and 4, and bending waves due to f-modes with
$l-m=1$, 3, and 5 \citep{Mankovich,French19}.

The detection of Saturn's oscillations opens a new window into the
internal properties of that planet.  
By comparing observed and theoretical f-mode frequencies, one can
learn about Saturn's background state.  For example, \cite{Mankovich}
inferred the bulk rotation rate with a precision of 10\%.
\citet{Fuller14} and \cite{Fulleretal14} showed that one of the observational
surprises--- specifically, that the density waves associated with
$m=2$ and $m=3$ prograde sectoral f-modes appear to be split into
multiple waves with slightly different frequencies---can be explained
by f-mode mixing with g-modes. This provides evidence for stable
stratification near Saturn's core.

In this work, we are concerned with a different issue.  The observed
ring waves have dimensionless amplitudes that range from a few percent
to unity. These, when translated into f-mode amplitudes, imply surface
displacements all within an order of magnitude of $1$ meter (see Table
\ref{tab:num} below). This appears to be rather fortuitous: if the
mode amplitudes were much less than a meter, the ring waves will be
too weak to be visible.
%
Why is Saturn oscillating at the amplitudes that we observe? And in
tandem, what excites these oscillations in Saturn?

Oscillations in stars are known to be driven by linear instabilities
(e.g., the $\kappa$-instability as for Cepheids and RR Lyraes), or
stochastic processes (e.g., turbulent convection for the Sun and red
giants). Saturn's oscillations must belong to the latter
category. Mode amplitudes are so small (dimensionless amplitudes $\sim
10^{-9}$) that their damping must be linear.  Hence if their
excitation was linear too, the mode amplitudes would grow (or shrink)
indefinitely.  Stochastic driving, on the other hand, can lead to a
variety of outcomes. Consider the example of throwing pebbles at a
bell. The pebbles come at random intervals and excite the bell's
oscillation incoherently.
Given a long enough time, the bell will settle into the so-called
fluctuation-dissipation equilibrium, with a ringing volume that
depends on, among other things, how hard a pebble is thrown, how many
pebbles are thrown per unit time, and how quickly the bell loses
energy.




What stochastic process could be operating inside a giant planet?
Much discussion in the past has focused on turbulent
convection. Although Saturn is likely fully convective, we argue here
that its internal flux is too low, and convection too feeble, to
explain the observations. Instead, we propose that past impacts drive
these modes to the observed amplitudes.

Stimulated by the comet Shoemaker-Levy 9's impact onto Jupiter,
\citet{Kanamori93,Marley94,Lognonne94,DB95} considered how impacts
affect internal oscillations \citep[also see][]{MarkhamStevenson}.
But these studies only considered small bodies like SL9, which has a
radius of $R \sim 2 \km$. They failed to consider the role of an
additional axis, that of time.

If the f-modes are very weakly damped, they can have a long memory of
past impacts, including by bodies much larger than the SL9
comet. Here, we explore this direction by first studying how weakly
damped the modes are (\S \ref{sec:densitywave},\ref{sec:damping}),
finding that the ring waves, as opposed to internal dissipation in
Saturn, dominate the damping.  We then expose the failure of
previously proposed mechanisms in exciting the f-modes to their
observed amplitudes (\S \ref{sec:otherdriving}), before turning to
consider how past large impacts can contribute (\S
\ref{sec:bombardment}).  Having calculated in detail the case for
Saturn, we extrapolate our calculations to Jupiter, and find that
there are observable consequences (\S \ref{sec:jupiter}).

Much of the discussions in this work are order-of-magnitude in nature,
as some physics is difficult to model in detail.

\section{Preparation: f-modes in Saturn}
\label{sec:prep}

We assemble some scalings for f-modes that are of use throughout this
paper.  The asymptotic expression for f-mode frequency as a function
of spherical degree $\ell$ is
\begin{equation}
\omega^2 = {{G M_s}\over{R_s^3}} \sqrt{\ell(\ell+1)}\, ,
\label{eq:omega}
\end{equation}
where $M_s, R_s$ are Saturn's mass and radius, respectively.  This
expression coincides fairly well with results from eigenmode
calculations, under the Cowling approximation \citep[also see][without
Cowling]{Mankovich}.  There is no $\ell=1$ f-mode -- although the
Cowling approximation (as we adopt here) may give it a non-zero
frequency, this mode should have zero frequency in the full solution
\citep{Christensen}.

We ignore the effects of planet rotation on the eigenfunction and
split the eigenfunction into angular and radial dependencies as in
eq. \refnew{eq:xirylm}.  The displacement vector is $\boldxi$, and we
define a normalized displacement vector ${\tilde\boldxi}$ that
satisfies
\begin{equation}
\int d^3 r \rho {\tilde \boldxi} \cdot {\tilde \boldxi} = M_s R_s^2\, ,
\label{eq:mynorm}
\end{equation}
which implies that
${\tilde\xi}$ has the dimension of length.  We focus on the radial
dependency of the eigenfunction here. 
For f-modes, the surface radial
displacement satisfies the following scaling,
\begin{equation}
{\tilde \xi}_r (r=R_s) \approx 1.8 R_s \times \sqrt{\ell (\ell+1)}\,  .
\label{eq:xirnorm}
\end{equation}
This rises with $\ell$ because high-$\ell$ f-modes are more
concentrated towards the surface. The physical radial
displacement is obtained by multiplying the above value by 
$Y_{\ell m}(\theta,\phi)$. 

Energy stored in a f-mode (including gravitational, compressional and
kinetic) is
\begin{equation}
E_{\rm mode} = {1\over 2} \omega^2 \int d^3 r \rho \boldxi \cdot \boldxi\, .
\label{eq:defE}
\end{equation}
With eqs. \refnew{eq:omega} and \refnew{eq:xirnorm}, this yields
\begin{eqnarray}
E_{\rm mode} & = & 
{{\sqrt{\ell(\ell+1)}}\over 2} {{GM_s^2}\over{R_s}} 
\left({{\xi_r}\over{\tilde \xi_r}}\right)^2 \nonumber \\
 & \approx & {{2\times 10^{26}}\over{\sqrt{\ell(\ell+1)}}}
 \erg \times \left({{\xi_r}\over{1 \m}}\right)^2\, ,
\label{eq:Escale}
\end{eqnarray}
where $\xi_r$ is that  evaluated at the planet surface.

Density perturbations by an f-mode produce perturbations in the
gravitational potential.  For a Eulerian density perturbation of the form
$\rho'(\bold{r})=\rho'(r)Y_{lm}(\theta,\phi)$, the perturbed potential outside
of Saturn is
\begin{equation}
 \Phi_{lm}(\bold{r})=-4\pi {GM_s\over R_s} {1\over 2l+1}{R_s^{l+1}\over r^{l+1}}Y_{lm}(\theta,\phi)J_{lm}
\label{eq:phi}
\end{equation}
where $J_{\ell m}$ is the gravitational moment
\citep{BinneyTremaine,MP93} and
\begin{eqnarray}
  J_{\ell m} &=& {1\over{M_s R_s^\ell}} 
\int_0^{R_s} r'^\ell \rho' (r') Y_{\ell m}(\theta',\phi') Y_{\ell m}^*(\theta',\phi')\, d^3
r' \nonumber \\
& = &{1\over{M_s R_s^\ell}}  \int_0^{R_s} r'^\ell \rho'(r') r'^2 dr' \, .
\label{eq:JL}
\end{eqnarray}
An orbiter, be that a particle in Saturn's ring, or a man-made
spacecraft orbiting the planet, can measure this potential variation.
Numerically, this can be
adequately approximated as
\begin{equation}
J_{\ell m} \sim 0.6 \left({{\xi_r}\over{\tilde \xi_r}}\right) \sim {{5.8\times
10^{-9}}\over{\sqrt{\ell(\ell+1)}}} \left({{\xi_r}\over{1\m}}\right)\, .
\label{eq:JL2} 
\end{equation}
for all $\ell$-values of practical interest.

\section{F-modes are damped by the Rings}
\label{sec:densitywave}

Saturn's f-modes are visible thanks to its sensitive rings. In the
following, we first show that Saturn's rings is a leading player in
removing energies from the normal oscillations, via the very density
waves that reveal their presence.  We then estimate the corresponding
f-mode energies, based on the observed amplitudes of these density
waves.  Both of these results set stringent constraints on possible
excitation mechanisms for the f-modes.

\subsection{Damping time}

We calculate the timescale on which density waves in the C-ring damp f-modes:
\be
t_{\rm damp}\equiv {{E_{\rm mode}}\over{\dot{E}_{\rm wave}}}\, ,
\label{eq:tdamp}
\ee
where $\dot{E}_{\rm wave}$ is energy input rate into
density waves and $E_{\rm mode}$ is the mode energy in Saturn.  The
former is $\dot{E}_{\rm wave}=\omega_p\Gamma$, where $\omega_p$ is the
angular pattern frequency of the wave (defined below) and $\Gamma$ is
given by the standard torque formula \citep{GoldreichTremaine78}: 
\begin{equation}
\Gamma = -m\pi^2{\Sigma\over rdD/dr}\Psi_{lm}^2 \, ,
\label{eq:torque}
\end{equation}
where
\begin{equation}
\Psi_{lm}\equiv {rd\Phi_{lm}\over dr}+{2\Omega\over
  \Omega-\omega_p}\Phi_{lm} \, .
\label{eq:psi}
\end{equation}
Here, 
 $\Sigma$ is the ring's surface density, $\Omega = \Omega(r)
$ is the Keplerian frequency at distance $r$, and
all quantities are to be evaluated at the Lindblad
resonance, i.e., where
 $D\equiv
\Omega^2-m^2(\Omega-\omega_p)^2=0$.\footnote{For simplicity, we
  have neglected the contribution due to the rotational flattening of
  Saturn.}
 For the special case of outer Lindblad resonances of
sectoral modes ($|m|=l$), the potential (eq. \ref{eq:phi}) evaluated
in the ring plane yields
\begin{eqnarray}
 |\Psi_{ll}| &=&
(3l+1)|\Phi_{ll}|
 \nonumber \\
&=&\left[ 4\pi {3l+1\over 2l+1} 
\left( {2l+1\over 4\pi}{1\over (2l)!} \right)^{1/2}  {(2l)!\over 2^ll!} {.84\over (l(l+1))^{1/4}}\right] \nonumber \\
& & 
\times
{GM_s\over R_s}{R_s^{l+1}\over r^{l+1}}  \left( {E_{\rm mode}\over GM_s^2/R_s} \right)^{1/2}
\label{eq:psi} 
\end{eqnarray}
after using Equations (\ref{eq:Escale}) and (\ref{eq:JL2}) to relate
$J_{ll}$ to $E_{\rm mode}$.
Therefore,
\begin{eqnarray}
{1\over t_{\rm damp}}
&=&{{\omega_p\Gamma}\over E_{\rm mode}} 
= {1\over \omega_p} \pi^2{\Sigma(l+1)\over 3l}|\Psi_{ll}|^2 {1\over
  E_{\rm mode}}
\nonumber \\
&= &{G\Sigma\over\omega_pR_s}
\left({R_s^{l+1}\over r^{l+1}}  \right)^2 \nonumber \\
& & \times \left[{0.7\cdot
  4\pi^3(l+1)^{1/2}(3l+1)^2(2l)!
\over 3 l^{3/2}(2l+1)2^{2l}(l!)^2}\right]\, .
\label{eq:tdamp}
\end{eqnarray}

\begin{table*}[ht]
\caption{F-mode properties, as inferred from ring seismology}
\centering
\begin{tabular}{c | c c c  | c c c c c}
\hline\hline\
$l=|m|$ & $\omega_p$ & $r_{\rm Lindblad}$ & $t_{\rm damp}$ 
& $\Sigma$ 
& $t_{\rm  damp}$ 
& $\Delta\Sigma/\Sigma$ 
&$E_{\rm mode}$ & $\xi_r$
\\
units: & $[10^{-4}$s$^{-1}]$  &$[R_s]$ &  $[\Sigma_1^{-1}$Myr] 
& $[\g/\cm^2]$ &  [Myr]& & $[10^{25}$ erg] & $[\m]$
\\
\hline
2 & 3.76 & 1.45 & 0.15 
& 4.0 &  0.037  & 0.47 & 0.28 & 0.18
\\
3 & 3.50 & 1.41 & 0.30 
& 6.9 & 0.044   &0.56          & 2.15  &0.61
\\ 
4 & 3.35 & 1.39 & 
0.25 & 5.8 & 0.10    &0.39           &  1.38  & 0.56
\\
5 & 3.22 & 1.39 & 1.27 & 
5 & 0.25    &0.12              &  0.21 & 0.24
\\
6 & 3.11 & 1.40 & 2.83 
& 5 & 0.57      & 0.056            & 0.10 & 0.18
\\
7 & 3.01 & 1.41 & 6.52 
& 5 &  1.3      &0.082               & 0.52  &0.44
\\
8 & 2.94 & 1.42 & 
15.4
& 5 & 3.1      &  0.050              &0.47 & 0.45
\\
9 & 2.87 & 1.43 & 37.0 
& 5 &7.4      & 0.099           &4.5 & 1.46
\\
10 & 2.82 & 1.44 & 90.3 
 & 4.95 &18.2     & 0.44     &218 & 10.7\\
\hline
\end{tabular}
\label{tab:num}
\end{table*}

We list values of $t_{\rm damp}$ in the fourth column of Table
\ref{tab:num} for a ring of surface density $\Sigma = 1\g/\cm^2$ (or
$\Sigma_1 = \Sigma/(1\g/\cm^2)= 1$), taking values of $\omega_p$ and
locations of resonances from Table 2 of \citet{HN18}.  Although there
are multiple waves associated with the $l=2$ and 3 f-modes, for
simplicity we only list the waves from their Table 2 that have the
highest amplitudes.  Including the other waves only changes our
results by a factor of a few.  The fifth column of Table \ref{tab:num}
lists values of $\Sigma$, ring surface density, as obtained from Table
5 of \citet{HN14}, which are in turn inferred from the wavelengths of
the density waves\citep{baillieetal11}; for waves not included in the
table of \citet{HN14}, we simply use $\Sigma=5 \g/\cm^2$.  The
realistic damping times, now accounting for the surface densities, are
listed in column 6 of Table \ref{tab:num}, and plotted in
Fig. \ref{fig:fmode-ring}. They increase with $\ell$, from a few
$10^4$ yrs to over $10^7$ yrs. So while the low-$\ell$ modes are
quickly dissipated, high-$\ell$ modes can retain energies for almost
as long as the putative age of the rings, $\sim 100$ Myrs as estimated
by \citet{Zhang}.  An adequate but crude fit is $t_{\rm damp} \propto
[\ell(\ell+1)]^2$.

It is conventional to define an oscillation quality factor, which is
the energy loss over one oscillation period,
\begin{equation}
Q \equiv {{E_{\rm mode}}\over{\oint\,  {{dE}\over{dt}}\, dt}}  \, .
\label{eq:defineQ}
\end{equation}
So $Q = 2\pi t_{\rm damp}/\omega$. Observed f-modes are damped by the
ring with $Q = 10^8$ to $Q=10^{11}$ (Fig. \ref{fig:fmode-ring}).


\subsection{Amplitude of f-mode}

\begin{figure}
        \centering
   \includegraphics[width=0.48\textwidth,trim=10 170 20 120,clip=]{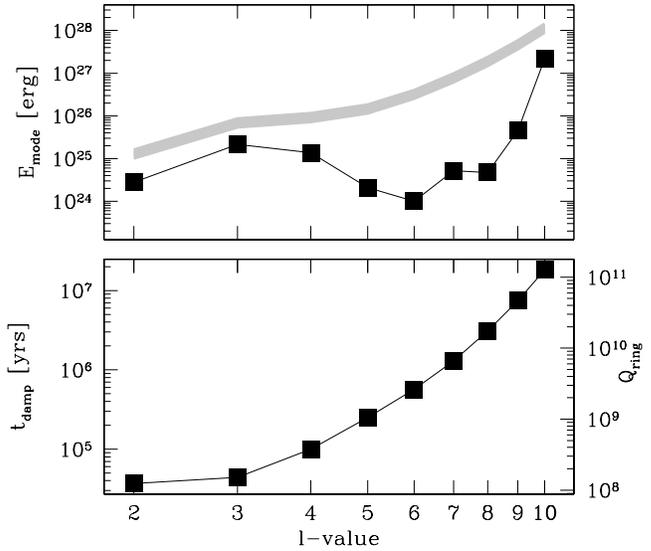}
   \caption{F-modes and the rings.  The top panel shows the mode
     energies as inferred from ring seismology, as a function of
     spherical degree (also see Table \ref{tab:num}). The grey curve
     indicates the energy value at which the ring density wave will
     have reached unity amplitude.  The bottom panel shows the damping
     time-scale for f-modes as they excite density waves in the
     C-ring, with the right axis approximately marking the
     corresponding quality factor. By contrast, internal damping
     typically give $Q \geq 10^{14}$.}
    \label{fig:fmode-ring}
\end{figure}

We now use the observed density waves to infer the amplitudes of
f-modes inside Saturn.
The amplitude of a C-ring wave at its first peak is \citep{Sicardy,MP93}
\be
{\Delta\Sigma\over\Sigma}\approx |\Psi_{ll}|{1\over 2\pi \Sigma r G}\, .
\label{eq:dsig}
\ee 
Adopting values for $\Delta \Sigma/\Sigma$ from \citep{HN18}, we
obtain values for the mode energies using eq. \refnew{eq:psi}, and
amplitudes for the surface displacements using
eq. \refnew{eq:Escale}. These results are listed in Table
\ref{tab:num} and plotted in Fig. \ref{fig:fmode-ring}.


While amplitudes of the associated density waves ($\Delta
\Sigma/\Sigma$) have a spread of a factor of $10$, mode energies have a
much larger spread, ranging from $10^{24} \erg$ to $10^{27} \erg$,
with the $\ell=10$ mode having the highest energy.
There is a characteristic mode energy at which ring waves become unity
amplitude ($\Delta \Sigma/\Sigma = 1$). These are indicated in
Fig. \ref{fig:fmode-ring}. The $\ell=2, 3$ and possibly the $\ell=10$
modes are close to this limit. In fact, visual inspections of the
optical depth profile \citep{HN13} associated with the $\ell=2$ and
$3$ waves would argue that they are  above unity. Here, we
assume that the damping rates for the f-modes are not strongly affected by this
nonlinearity.

\section{Internal Damping Are Weak}
\label{sec:damping}

We consider various energy sinks for the f-modes inside Saturn. These
include mode damping by radiative diffusion (near the photosphere
where the thermal timescale is the shortest), by leakage of mechanical
energy (into the isothermal stratosphere above the photosphere), and
by viscosity associated with the turbulent convection (throughout the
planet). We show that all of these yield $Q \sim 10^{14}$ or higher,
and cannot compete with the above discussed ring
damping. Previous work has come to very different conclusions
regarding  internal damping
\citep[e.g.][]{MarkhamStevenson}. Because the correct damping is
important for estimating the level of stochastic excitation
(eq. \ref{eq:stochastic1}), we spend much labour in this
section. Readers who are more interested in the excitations can jump
ahead to \S \ref{sec:otherdriving}.

For an inviscid fluid, internal heat can be removed by heat diffusion and
mechanical work,
\begin{equation}
dQ = TdS - p d\left({1\over\rho}\right)\, .
\label{eq:thermodynamics}
\end{equation}
Energy in the pulsation is changed as described by, e.g., eq. (25.7) of \citet{UnnoBook},
\begin{eqnarray}
{{dE}\over{dt}} &=& {d\over{dt}}\int_0^M dM ({1\over 2} v^2 + {1\over 2} \Phi + Q)\nonumber \\
& = & \int_0^M dM\, {T {{dS}\over{dt}}  - \int_{R_s} p {\boldv} \cdot d {\bf S}}\, ,
\label{eq:planetE}
\end{eqnarray}
where the latter is a surface integral evaluated at Saturn's photosphere.
For a pulsation that is nearly strictly periodic and adiabatic,
i.e., a Lagrangian perturbation of a complex form $\delta p
=|\delta p| e^{i \theta_p} e^{i \omega t}$, with $\omega = \omega_r +
i \omega_i \approx \omega_r$ ($|\omega_i| \ll \omega_r$), over one
cycle, it gains energy
\begin{eqnarray}
&& \Delta E = \Delta E_{\rm rad} + \Delta E_{\rm leak} = \oint dt
{{dE}\over{dt}} = \nonumber \\
& &{ \pi\over\omega} \int_0^M dM\, {\rm Re}\left[\delta T^* {{d\delta s}\over{dt}} \right]
- 4 \pi^2 R_s^2  {\rm Im}\left[ \delta p^* \xi_r\right]_{R_s}\, ,
\label{eq:dEdt}
\end{eqnarray}
where quantities in the last expression have shed their oscillatory
time-dependencies after the time-integration.  The first term ($\Delta
E_{\rm rad}$) is due to damping by internal heat diffusion, and the
second term ($\Delta E_{\rm leak}$) is due to leakage of mechanical
energy to vacuum.

For a viscous fluid, the fluid equation of motion has an additional
friction term, 
\be
\rho \ddot{\boldxi} 
=- \bnabla p^\prime + {{\bnabla p}\over{\rho}}\, \rho^\prime - \rho
\bnabla \delta \Phi + {\mbox{\boldmath $F_{\nu}$}},
\label{eq:eqnmotion3}
\ee
where the viscous force
\be
{\mbox{\boldmath $F_{\nu}$}} = \bnabla \cdot \left(\rho \nu \bnabla
{\dot{\boldxi}}\right)\, .
\label{eq:definef}
\ee
In our case, the viscosity $\nu$ arises from turbulent convection. 
Over one cycle, the mode energy changes by
\be
\Delta E_{\rm vis}
=\oint dt \int d^3 r\, {\dot{\boldxi}} \cdot {\mbox{\boldmath $F_{\nu}$}}
= {\pi \omega {{\int d^3 r\, \rho \nu \,\boldnabla \boldxi : \boldnabla \boldxi}}}\, .
\label{eq:definegamma}
\ee

The most rigorous way to obtain the corresponding quality factor $Q$
(eq. \ref{eq:defineQ}) is to solve the non-adiabatic f-mode
eigenfunctions. In the following, we take the quicker path of
estimating the magnitudes of $\Delta E_{\rm vis}$, $\Delta E_{\rm
  rad}$ and $\Delta E_{\rm leak}$. This is 
a delicate procedure, but it is ultimately more physically illuminating.
We are able to show that the three dampings are all exceedingly weak
($Q \geq 10^{14}$).

For simplicity, we suppress the $\ell$-dependence of the f-modes as we
are concerned only with low-$\ell$ modes. We further assume that all
fractional Lagrangian perturbations are of the same order, $\delta
\rho/\rho \sim \delta T/T \sim \delta F_{\rm rad}/F_{\rm
  rad}$. 
Saturn's atmosphere is taken to be fully convective up to the
photosphere, above which there is an isothermal stratosphere. This
ignores possible stratification by the so-called 'moist
convection' \citep{Ingersoll95}, or the putative sub-adiabatic zone
as suggested by \citet{Guillot94}.


\subsection{Turbulent Damping}

The interior of Saturn is convectively unstable. Energy-bearing
turbulent eddies, driven by thermal buoyancy, have a scale of order
the local scale height, $H$, and a characteristic
velocity obtained by $\rho v_{\rm cv}^3 = F$, where $F$ is the
internal cooling flux and $F \approx 2200 \erg/\s/\cm^2$ for
Saturn. Eddies in the so-called inertial range are also present as the
energy-bearing eddies turn-over and cascade down in scale.  Taking the
kinematic viscosity of the form 
\be \nu \sim v_{\rm cv} H
\left[ {1\over{1+(\omega \tau_{\rm cv}/2\pi)^s}}\right]\, ,
\label{eq:nucv-main}
\ee with the factor in the square braket representing the reduction in
viscosity when the typical convection turn-over time ($\tau_{\rm cv}
=H/v_{\rm cv}$) is much longer than the mode period, and
inertial-range eddies, as opposed to the more powerful energy-bearing
eddies, dominate the viscous damping. Even at the photosphere where
convection is the fastest, adopting $H \sim 30\km$, $\rho_0 \sim
10^{-4} {\bar \rho}$, we find that $v_{\rm cv} \sim 300 \cm/\s $,
$\tau_{\rm cv} \sim 10^4 \s$, and $\omega \tau_{\rm cv} \gg 1$.  In
the following, we adopt $s=2$ as is appropriate for a Kolmogorov
turbulence spectrum \citep{GoldreichNicholson77}, and find that
$\omega \tau_{\rm cv} \gg 1$ throughout the planet and that damping is
dominated by eddies in the surface scale height, with $\nu \sim F/(\omega^2 \rho
H)$  \citep[see][for more details]{Wu05b}. So eq. \refnew{eq:definegamma} yields
\begin{equation}
\Delta E_{\rm vis} \sim 4 \pi^2 R_s^2 {F\over \omega}
\left({{\xi_r}\over R_s}\right)^2 
\sim  {{\pi L_{\rm umin}}\over\omega}
\left({{\xi_r}\over R_s}\right)^2 \, .
\label{eq:dEvis}
\end{equation}
where $L_{\rm umin} = 4 \pi R_s^2 F$ is Saturn's intrinsic
luminosity. 
Scalings in \S \ref{sec:prep} imply that $E_{\rm mode} \sim 0.5 \omega^2
M_s R_s^2 (\xi_r/{\tilde \xi_r})^2$, so
%
\begin{equation}
Q_{\rm vis} = {{E_{\rm mode}}\over{\Delta E_{\rm vis}}}  \sim 
 10^{14} \left( {{\omega}\over{10^{-3} \s^{-1}}}\right)^3 \, .
\label{eq:Qvis}
\end{equation}

\subsection{Radiative Diffusion}

Here, we consider heat diffusion by radiation between fluid parcels of
different temperatures.  In principle, convection can also thermally
respond to pulsation and leads to further heat diffusion. But we
ignore convection here because, first, there is no good first-principle
treatment of perturbed convective flux; and second, the convection turn-over
time is very long compared to the mode period and hence we may be justified
to ignore its perturbations. Most of the diffusive heat-loss comes
from the scale height just below the photosphere, where the diffusion
time is the shortest. This is also where the convective flux is
rapidly substituted by radiative flux, so $F = F_{\rm rad} + F_{\rm
  cv} \sim F_{\rm rad} \sim (4 a c T^4)/(3\kappa \rho) d\ln T/dr$, and
$dF_{\rm rad}/dr \sim F/H$, where $H$ is the local scale height.

The equation of energy conservation reads,
\begin{equation}
T {d{\delta s}\over{dt}} =  - \delta \left({1\over \rho} \boldnabla \cdot {\boldF}\right) \, .
\label{eq:dsdF}
\end{equation}
Ignoring the perturbation to the convective flux, we set $\delta F =
\delta F_{\rm rad}$ and expand the right-hand side in Lagrangian
variables as
\begin{eqnarray}
\delta \left({1\over \rho} \boldnabla \cdot {\boldF}\right) & = &
- {1\over \rho}  {{\delta \rho}\over{\rho}} \boldnabla \cdot {\boldF} 
+ {1\over \rho}  {{\delta \rho}\over{\rho}} \boldnabla \cdot {\boldF} \nonumber \\
& & + {1\over \rho} {d\over{dr}} \left({{\delta F_{\rm rad}}\over{F_{\rm rad}}}\right) F_{\rm rad} 
+{1\over \rho} {{\delta F_{\rm rad}}\over{F_{\rm rad}}} {{dF_{\rm rad}}\over{dr}} \nonumber \\
& \sim & {F \over {\rho H}}{{\delta F_{\rm rad}}\over F_{\rm rad}} \, .
\label{eq:dFF}
\end{eqnarray}
The term that involves $dF_{\rm rad}/dr$ dominates on the
right-hand-side, since for f-modes, $\delta F_{\rm rad}/F_{\rm rad} \sim \delta \rho/\rho$
varies much more smoothly (see Fig. \ref{fig:plot-mode}).
Most of the contribution to damping comes from the scale height
immediately below the photosphere, where the entropy perturbation is
the largest  (eq. \ref{eq:dsdF}). And we find
\begin{eqnarray}
\Delta E_{\rm rad} & = & {\pi\over{\omega}}
 \int_0^M dM\, {\rm Re}\left[ \delta T^* {{d\delta s}\over{dt}}\right] \nonumber \\
& = & {\pi \over \omega} \int_0^M dM\, {\rm Re}\left[ \left({{\delta T}\over T}\right)^* \delta ({1\over \rho} \boldnabla\cdot {\boldF})\right]\nonumber \\
& \sim & {{\pi L_{\rm umin}}\over{\omega}} \left({{\delta \rho}\over{\rho}}\right)^2\, ,
\label{eq:dErad}
\end{eqnarray}
Because f-modes are highly incompressible, $\delta \rho/\rho \ll
\xi_r/r$ (see Fig. \ref{fig:plot-mode}), this term is much smaller
than damping by turbulent viscosity (eq. \ref{eq:dEvis}).

 We quantify
the compressibility by
\begin{equation}
{\cal F} \equiv {{\left|{{\delta \rho}\over\rho}\right|}\over{\left|{\xi_r}\over{r}\right|}} \ll  1\, ,
\label{eq:definecomp}
\end{equation}
evaluated near the surface.  Following \citet{MarkhamStevenson}, we
introduce the radiation constant $\tau_{\rm rad} \approx c_p T/\kappa
F$.
This is of order $\tau_{\rm rad} \sim 10^7 \s$ at Saturn's
photosphere, defined as where $\kappa \rho_0 H \sim 2/3$.  Value for
the surface scale height is $H \sim 30$km, and the photospheric
density $\rho_0 \sim 10^{-4} {\bar \rho}$, with ${\bar \rho}$ being
the mean density of Saturn.  We can now recast our result in their
notation as
\begin{eqnarray}
  Q_{\rm rad} & = & {{E_{\rm mode}}\over{\Delta E_{\rm rad}}} \nonumber \\
  & \sim &
  (\omega \tau_{\rm rad}) \left({\omega\over{\omega_{\rm
          dyn}}}\right)^2 \left({{\bar \rho}\over{\rho_0}}\right)
  \left({{R_s}\over{H}}\right)^2 {\cal F}^{-2}\nonumber \\
  & \sim & 10^{13} \left( {{\omega}\over{10^{-3} \s^{-1}}}\right)^3
  {\cal F}^{-2}\, ,
\label{eq:Qrad}
\end{eqnarray}
where $\omega_{\rm dyn} = \sqrt{G {\bar \rho}} \sim g/R_s$.

\subsection{Wave Leakage}

This is the most subtle damping to estimate.  Material above the
photosphere also oscillates and can in principle communicate the wave
energy to infinity (or steepen into a shock at some height).
For adiabatic oscillations, there is exactly zero leakage for waves
with frequencies below the so-called acoustic cutoff frequency.
This is true for all f-modes of concern.  However, this is modified
when entropy perturbation is considered (non-adiabatic
perturbations). Our calculation below argues that $Q_{\rm leak} \gg
10^{13}$. In contrast, \citet{MarkhamStevenson} found that $Q_{\rm
  leak} \sim 10^7$.

We assume the layer above the photosphere to be isothermal, as
suggested by observation. So the scale height $H = \Gamma_1 p/g\rho =
-\Gamma_1 d\ln p/dr $ remains largely constant. Here, the adiabatic
index $\Gamma_1 = (\partial \ln p/\partial \ln \rho)_s$. We first
consider adiabatic perturbations before turning toward non-adiabaticity.

To obtain the wave characteristics at the photosphere and above, we
adopt the form of adiabatic perturbation equation as in eq. (8) of
\citet{GoldreichWu99},
\begin{equation}
{{d^2}\over{dr^2}} \left({{\delta p}\over p}\right) - {{\Gamma_1}\over{H}} {d\over{dr}} \left({{\delta p}\over p}\right) + \left[k_h^2 ({{N^2}\over{\omega^2}} - 1) + {{\omega^2}\over{c_s^2}}\right] = 0\, .
\label{eq:dpp}
\end{equation}
Here, $k_h^2 = \ell(\ell+1)/r^2$, the \Bruntfreq frequency $N^2 = g (d\ln \rho/dr -
1/\Gamma_1 d\ln p/dr) = (\Gamma_1 - 1) g/H $, and $c_s^2 = \Gamma_1
p/\rho = gH$. As all coefficients are constants on scale of $H$, the
solution has the form $\delta p/p \propto \exp(k_r r)$, where
\begin{equation}
k_r = {1\over 2}{{\Gamma_1}\over{H}} \pm {1\over 2} \sqrt{ \left({{\Gamma_1}\over H}\right)^2 - 
4 \left[ k_h^2 ({{N^2}\over{\omega^2}} - 1) + {{\omega^2}\over{c_s^2}} \right]}\, .
\label{eq:krsolution}
\end{equation}
The term in the square bracket is negative when $\omega \geq
\omega_{\rm ac} = \Gamma_1 c_s/2H$, where $\omega_{\rm ac}$ is the
so-called acoustic cut-off frequency.\footnote{It is also negative
  when $\omega \leq (H/R) \omega_{\rm ac}$, when the propagating wave
  is g-mode in nature. This is too low frequency to be of relevance.}
In this case, the mode is propagative into the isothermal layer and it
is strongly damped by wave leakage.  On the other hand, when $\omega <
\omega_{\rm ac} \sim 0.017 \s^{-1}$ (or a period of $370\s$), as is
the case of all f-modes we are concerned with,
propagation is forbidden and the wave is evanescent in the isothermal
layer. All energy flux is reflected back to and contained within the
planet \citep{UnnoBook}.

For the latter waves, causality dictates that one has to take the
negative sign branch in eq. \refnew{eq:krsolution}, because otherwise
the energy density ($\rho \xi_r^2$) will rise exponentially with
height. This leads to
\begin{equation}
k_r \approx {{k_h^2 ({{N^2}\over{\omega^2}} - 1) + {{\omega^2}\over{c_s^2}}}\over{ \Gamma_1 /H}}  \sim {\cal O}({1\over R_s})\, ,
\label{eq:krtrue}
\end{equation}
or that fluid displacements, as well as Lagrangian perturbations, can
only vary on the scale of the planet radius. The nearly constant $\xi_r$,
for instance, indicates both that the whole atmosphere is lifted up in
unison, and that the energy density $\rho \xi_r^2$ decays
exponentially with height. So the isothermal atmosphere reflects all
energy density back into the interior.
In other words, since the atmosphere communicates across with the
speed of sound, for perturbations that vary more slowly than the acoustic
cut-off, the atmosphere remains hydrostatic and cannot provide
a restoring force.

The above conclusion is modified when radiative diffusion
interferes. The perturbations, now no longer adiabatic, experience a
phase shift between $\delta p/p$ and $\xi_r$. This
introduces wave leakage. We estimate the magnitude of this effect.

Let the photospheric pressure be $p_0$,
\begin{equation}
{\rm Im} \left[\xi_r^* \delta p\right]_{R_s} \approx R_s p_0\left. \left({{\xi_r}\over {r}} {{\delta p}\over p}\right)\right|_{R_s} \times (\delta \theta)\, ,
\label{eq:xirdp}
\end{equation}
where the difference in phases between the complex $\delta p/p$ and
$\xi_r$, $\delta \theta$, is caused by changes in entropy,
\begin{equation}
- \boldnabla \cdot \boldxi =  {{\delta \rho}\over{\rho}} = {1\over{\Gamma_1}} {{\delta p}\over p} + 
\rho_s \delta s\, .
\label{eq:drhods}
\end{equation}
Here $\rho_s = (\partial \ln \rho/\partial s)_p$. The entropy
perturbations, in term, is related to flux retention
(eq. \ref{eq:dsdF}), which, for an isothermal atmosphere where
$dF_{\rm rad}/dr = 0$ applies, is dominated by the third term on the
right-hand-side of eq. \refnew{eq:dFF}. This produces
\begin{equation}
i \omega T \delta s \sim {1\over \rho}{d\over{dr}} \left({{\delta
      F_{\rm rad}}\over{F_{\rm rad}}}\right) \sim k_r {F\over \rho}
{{\delta \rho}\over{\rho}} 
\sim {F \over{R_s \rho}} {{\delta \rho}\over \rho} \, .
\label{eq:ds2}
\end{equation}
Compared to the magnitude of $\delta s$ in the underlying convection
zone, this value is smaller by a factor of $H/R_s$. This difference
arises because of the isothermal structure of the atmosphere, as well
as eq. \refnew{eq:krtrue}. This leads to a rough estimate for the
phase shift as
\begin{equation}
\delta \theta \sim {H\over{R_s}} {1 \over{\omega \tau_{\rm rad}}}\, ,
\label{eq:theta}
\end{equation} 
And an order of magnitude estimate for $Q_{\rm leak}$ is
\begin{eqnarray}
Q_{\rm leak} & \approx &
(\omega \tau_{\rm rad}) \left({\omega\over{\omega_{\rm dyn}}}\right)^2
\left({{\bar \rho}\over{\rho_0}}\right) \left({R_s\over H}\right)^2
{\cal F}^{-2}\nonumber \\
& \sim & 10^{13} \left( {{\omega}\over{10^{-3} \s^{-1}}}\right)^3
{\cal F}^{-2}\, .
\label{eq:Qleakvalue}
\end{eqnarray}
This, interestingly, is of the same order of magnitude as that by
radiative diffusion (eq. \ref{eq:Qrad}).  It reflects the fact that
both dampings scale with the internal flux, and both occur within a
scale height of the photosphere. Both of these damping are much weaker
than that by turbulent convection (eq. \ref{eq:Qvis}), because f-modes
are very incompressible (${\cal F} \ll 1$).



\subsection{Difference from \citet{MarkhamStevenson}}

\citet{MarkhamStevenson} considered the same physics of wave
leakage. But their approach led them to conclude a
$Q_{\rm leak}$ of the form (their eq. 61) 
\begin{eqnarray}
Q_{\rm leak,MS} 
& \approx &
(\omega \tau_{\rm rad}) \left({\omega\over{\omega_{\rm dyn}}}\right)^2 \left({{\bar \rho}\over{\rho_0}}\right) \nonumber \\
& \sim & 10^7 \left( {{\omega}\over{10^{-3} \s^{-1}}}\right)^3 \, .
\label{eq:wrongQ}
\end{eqnarray}
This implies a dissipation orders of magnitude stronger than our
estimate in eq. \refnew{eq:Qleakvalue}, with the difference being
$(R_s/H)^2 {\cal F}^{-2} \gg 10^{6}$. There are three reasons for this
difference. First, while considering the Lagrangian pressure
perturbation (their eq. 52), they equate $\delta p/p$ to $T'/T$, where
$T'$ is the Eulerian temperature perturbation, while we equate $\delta
p/p$ to $\delta T/T$.  Because of the large temperature gradient near
the surface (variation scale $H \ll R_s$), $T'/T = \delta T/T - \xi_r
dT/dr \approx (R_s/H) \delta T/T $, this brings about the first factor
of $R_s/H$ in the difference.  Second, we find that the wave is
evanescent in the isothermal region. This leads us to adopt an entropy
variation that is smaller than that in the convection zone below by a
factor of $H/R_s$ (see discussion around eq. \ref{eq:ds2}).  Lastly,
we argue that the incompressibility of f-modes should introduce a
factor of ${\cal F}^{-2}$ into the calculation of $Q$.

\section{Mode excitation: Convection and Storms}
\label{sec:otherdriving}

In this and the next sections, we will discuss plausible excitation
mechanisms. As the modes appear to be stalled at amplitudes that are
very weak, the most likely driving is stochastic in nature.  In the
presence of an external damping, stochastic forcing increases mode
energies linearly in time, capped by the equipartition value ($E_k$),
\begin{equation}
E_{\rm mode} = {\rm Max}\left\{ N \Delta E_{\rm kick}, E_{\rm
    k}\right\} \, ,
\label{eq:stochastic1}
\end{equation}
Here $N$ is the number of discreet kicks received within the mode's
damping time, $\Delta E_{\rm kick}$ the typical energy given to the
mode during one stochastic event, and $E_k$ the total energy of such
an event. The capping reflects the fact that while the event can
excite a mode, it can also absorb energy from the mode. According to
the fluctuation-dissipation theorem, equipartition is the best
possible outcome.  Typically, $\Delta E_{\rm kick} \ll E_k$. 

We argue above that the main damping for f-modes is via density waves
in the rings, as opposed to any other dissipation internal to Saturn.
This provides the requisite damping timescales.  Here, we first focus
on two previously proposed excitation mechanisms.

\subsection{Convection}

While there have been claims in the literature that turbulent
convection can excite Saturn's f-modes to the observed amplitudes
\citep{Marley91,MP93,Fuller14,Mankovich}, this is in fact
incorrect.

Due to the small intrinsic flux of Saturn, turbulent convection is
far too weak (in velocity) and too slow (in turn-over time)
to account for the observed mode energies---even in the absence of ring damping.

Stochastic excitation by fluctuating convective eddies occurs at a rate
\citep{GoldreichKumar,Goldreich94,MarkhamStevenson}
\begin{equation}
\left. {{dE}\over{dt}}\right|_{\rm cv} \sim 2 \pi \omega^2 \int_0^{R_s} dr \, r^2 \rho^2
\left|{{d {\tilde\xi_r}}\over{dr}}\right|^2 \int_0^{h_{\omega}(r)} {{dh}\over
  h} v_h^3 h^4\, ,
\label{eq:cvrate}
\end{equation}
where $v_h$ and $h$ are the eddy velocity and size, respectively.
Only eddies that turn-over faster than the mode period contributes,
or, $h \leq h_{\omega}$ where $h_{\omega}$ is the largest eddies for
which $(h/v_h) \sim 1/\omega$. These eddies are all in the inertial
range, since the energy-bearing eddies satisfy $\omega \tau_{\rm cv}=
\omega H/v_{\rm cv} \gg 1$, everywhere inside Saturn.  For a
Kolmogorov turbulent spectrum, $v_h \sim (h/H)^{1/3} v_{\rm cv}$,
so we only need be concerned with $h \sim h_\omega$ and the stochastic
driving rate is
\begin{equation}
\left.{{dE}\over{dt}}\right|_{\rm cv} \sim 2 \pi \omega^2 \int_0^{R_s} dr \, r^2 \rho^2
\left|{{d {\tilde \xi_r}}\over{dr}}\right|^2 v_\omega^3 h_\omega^4\, ,
\label{eq:cvrate2}
\end{equation}
where the velocity at $h_\omega$ is $v_\omega$ and we have $v_\omega =
\omega h_\omega$, and $h_\omega = H \left(1/\omega \tau_{\rm
    cv}\right)^{3/2}$. This integral is dominated by the contribution from
the top scale height. 

In the mean time, turbulence removes energy from the modes in the form
of turbulent viscosity (eq. \ref{eq:definegamma}). The kinematic
viscosity in eq. \refnew{eq:nucv-main} can be manipulated to become
the following form
\begin{equation}
\nu \sim v_\omega h_\omega\, .
\label{eq:nunew}
\end{equation}
Or, viscous damping is  dominated by the same eddies that drive the
modes. 

When the stochastic driving and damping are balanced, we reach 
the so-called 'fluctuation-dissipation' equilibrium, or 'energy equipartition'.
Both driving and damping are dominated by turbulent eddies 
near the surface, since $\omega \tau_{\rm cv}$ is
the smallest there and the relevant eddies ($h_\omega$) lie closest to the
energy-bearing eddies. 
Equating the driving in eq. \refnew{eq:cvrate2} with the damping from
eq. \refnew{eq:dEvis}, we find that mode energy is
equilibrated to that of the relevant eddy,
\begin{equation}
E_{\rm mode} \sim \rho v_\omega^2 h_\omega^3\, .
\label{eq:equipartition}
\end{equation}
Evaluating near the surface, we find 
\begin{equation}
E_{\rm mode} \sim 5\times 10^{14} \erg \left({{\omega}\over{10^{-3}
      \s^{-1}}}\right)^{-11/2} \, .
\label{eq:Ecv}
\end{equation}

Due to the steep dependence of the above expression on frequency, any
missing factors of $2$ or $\pi$ can dramatically affect the result. To
be be cautious, we calibrate our results against that obtained for the
Sun.  The most highly excited modes in the Sun are observed to have
energies of order $10^{28}$ ergs, at frequencies of $3$ mHz, or
$\omega \sim 0.02 \s^{-1}$.
Equation \refnew{eq:equipartition} implies a physical scaling
\begin{equation}
E_{\rm mode} \propto {{F^{5/2}}\over{\rho^{3/2} H^{5/2}
    \omega^{11/2}}}\, .
\label{eq:Escaling}
\end{equation}
The solar photosphere has $\rho \sim 2\times 10^{-6} \g/\cm^3$, scale height
$H \sim 30\km$, and  flux $F = 7\times 10^{10} \erg/\s/\cm^2$. Substituting
these into the above equation, we find that $E_{\rm mode}$ in Saturn
should be some $10^{-15}$ smaller than that in the Sun. This confirms
our result in eq. \refnew{eq:Ecv}.
Moreover, the predicted $E_{\rm mode}$ drops sharply with mode
frequency, as is also observed for solar p-modes with frequencies
above $3$ mHz.

Eq. \refnew{eq:Ecv} falls below by orders of magnitudes compared to
that found by \citet{MarkhamStevenson}. They concluded that, for
Jupiter, f-mode energies by turbulent convection should be only $3$
orders smaller than those in the Sun. Unfortunately, we fail to
reconstruct their results and so could not resolve our differences.

Eq. \refnew{eq:Ecv}  lies much below the f-mode energies inferred
from ring seismology (Table \ref{tab:num}). This excludes turbulent
convection as a plausible source of excitation. The situation is
further exacerbated by  the fact that Saturn's rings
damp the f-modes much more strongly than  internal dissipation. 
Eq. \refnew{eq:stochastic1} suggests that in this case, 
\begin{equation}
E_{\rm mode} \sim  5\times 10^{14} \erg \left({{\omega}\over{10^{-3}
      \s^{-1}}}\right)^{-11/2} \left({{Q_{\rm ring}}\over{Q_{\rm vis}}}\right)\, .
\label{eq:Emodesmall}
\end{equation}

\subsection{Water Storms, Rock Storms...}

In the presence of trace volatiles, convection can take a more
complicated form. Other than the quasi-equilibrium turbulent
convection studied above, there could be episodic, triggered
convection that is driven not by the small super-adiabatic gradient,
but by the latent heat release from the condensing volatile.
Such ``storm'' cells are observed on the surface of Jupiter
\citep{Gierasch00} and Saturn \citep{Sayanagi}, and when they occur,
can carry a fair fraction of the internal flux, ranging from a few
percent on Saturn \citep{LiJiang}, to $\sim 60\%$ on Jupiter
\citep{Gierasch00}.
Deeper storms ('rock storms'), possibly driven by the condensation of
refractory material (silicate, magnesium, iron-bearing) has also been
hypothesized to exist much deeper inside the planet (up to $10^4$
bars). 
%
\citet{MarkhamStevenson} considered whether these storms can excite
f-modes in Jupiter\citep[also see][]{Dederick}. They argued that
shallow water storms are too weak to be relevant, but introduced the
ingenious idea of a rock storm.  We re-evaluate their hypothesis here.

Updraft of a warm, moist parcel (water or silica) can hit into the
cloud layer on top, depositing its momentum and energy.  Analogous to
the effects of an external impact, this can excite f-modes (see section
below).  While storms on Earth, Jupiter, and Saturn are observed to
have a horizontal extent of order thousands of kms, here we assume
that only a spherical patch of order the local scale height can act as
an impulse in driving the modes.  Time coherence across different such
patches is likely weak and can be considered as random, uncorrelated
punches \citep{MarkhamStevenson}.

Following \citep{MarkhamStevenson}, relevant physical parameters for
the water storms and the (hypothetical) rock storms are listed in
Table \ref{tab:storm}.  In particular, the storm energy is
\begin{equation}
E_k \sim \rho_0 H^3 v_{\rm storm}^2\, ,
\label{eq:Ekstorm}
\end{equation}
where the storm velocity is related to the buoyancy acceleration
$v_{\rm storm}^2 \sim (f L_v/c_p T) g H$, with $L_v$ being the latent
heat for the volatile and $f$ the volatile mass fraction.  We adopt
the latter to be about $10$ times solar, for both the water and the
rock cases.
As Table \ref{tab:storm} shows, these storm cells contain orders
of magnitude more energy than that in a quasi-equilibrium turbulent cell
that has the same frequency as an f-mode (eq. \ref{eq:Ecv}).

\begin{table}
\caption{Storms and f-mode excitation}
\centering
\begin{tabular}{l | c c}
\hline\hline
 & 'ice storm' & 'rock storm' \\
\hline
depth ($H$) &  $30\km$ & $10^8\cm$\\
density ($\rho_0$) &  $10^{-4} {\bar \rho}$& $0.05 {\bar \rho}$ \\
latent heat & $2.3 \times 10^{10} \erg/g$ & $1.2\times 10^{11}
\erg/g$\\
volatile mass fraction & $5\%$ & $1\%$\\
buoyancy speed ($v_{\rm storm}$) & $10^4\cm/\s$ & $3\times 10^4 \cm/\s$ \\
energy in one storm cell ($E_k$) &  ${{10^{24}}}${ergs} & $10^{32} \erg$s\\
one-kick energy ($\Delta E_{\rm kick}$, $\ell=2$) & $10^{6} \erg$s &
${10^{22}}${ergs} \\
\hline
\end{tabular}
\label{tab:storm}
\end{table}

Let us take the simplistic picture that the condition conducive for the
storm builds up gradually, and then within a time short compared to the
f-mode period the entire storm cell releases the latent heat
contained within.\footnote{Storm momentum, released as the updraft
  hits into the cloud deck on top, cannot drive pulsation. Within the
  planet, momentum conservation requires that the down-ward momentum,
  equal in magnitude and opposite in sign, is also imparted at the
  same time.}
A shallow water storm has energy comparable to that of an $R=30\m$
impactor, while the deep rock storm is equivalent to that of an
$R=15\km$ impactor.  Using algebra similar to that for an meteor
impact (eq. \ref{eq:amplitude12}), we find that the one-kick energy,
$\Delta E_{\rm kick}$, for the $\ell=2$ f-mode, is $10^{6}$
ergs for the water storm and $10^{22}$ ergs for the rock storm, These
values grow with the mode's spherical degree as $\ell(\ell+1)$.


We can exclude water storms as an important source of
driving. Requiring that the $\ell=2$ mode be excited to its observed
value, $E_{\rm mode} = 3\times 10^{24} \erg$, within its damping time
of $\sim 4\times 10^4$yrs, one finds that the water storms need to
occur very frequently and transport a flux that is some $10^6$ times
higher than Saturn's total flux. In addition, energies in the
higher-$\ell$ modes actually exceed the equipartition
value.\footnote{\citet{Dederick} obtained, under some parametrization
  for the water storms, energies much larger than equipartition.  We
  argue that equipartition is the upper limit: if the storm can pump
  energy on the mode, the reverse process is also working to remove
  energy from the mode.}

The rock storms, on the other hand, are less conclusive.  The same
exercise for the $\ell=2$ mode shows that they only have to carry some
$2\%$ of the internal flux, and occur once every century.  At the
moment, we do not know if such deep storms exist, let alone their
frequency. So such a scenario cannot be excluded, or substantiated.

\section{Mode excitation: Impacts}
\label{sec:bombardment}

F-modes in Saturn have life-times that range from $4\times 10^4$ yrs
to $2\times 10^7$ yrs. Could large impacts that occur over these timescales
be responsible for exciting the f-modes to their current
amplitudes?  In other words, could Saturn still be trembling from
terrors of the deep past?

\subsection{Bombardment Rates on Saturn}

We first summarize observations on direct impacts onto Saturn.
\citet{SaturnRate} compiled impact rates on outer solar system bodies,
as a function of impactor size. These are based on impact crater
counts on the moons of the giant planets, and on the observed
populations of Kuiper Belt objects. For Saturn, they found that an
impactor of radius $R = 1\km$ arrives once every $\sim 100$ yrs, while
$R=100\km$ bodies arrive once every $\sim 10^8$ yrs. Impact rates in
their Fig. 2 can be roughly fit by a power-law,
\begin{equation}
\gamma_{imp} (R) \approx 10^{-2} \left({{R \over{1
        \km}}}\right)^{-2.8} {\rm \, yr}^{-1}\, .
\label{eq:brate}
\end{equation}
Such a scaling implies that the mass flux (and energy flux)  is
roughly flat with impactor size. Uncertainties in these
rates are listed at about an order of magnitude. Moreover, the size
distribution of Kuiper Belt objects is not a simple power-law but
contains structures. So this fit is a simplification. 

Jupiter, being more massive, sustains a bombardment rate some $2.5$
times larger.




\subsection{Energy Imparted to Normal Modes}

\begin{figure}
        \centering
   \includegraphics[width=0.48\textwidth,trim=10 150 40 120,clip=]{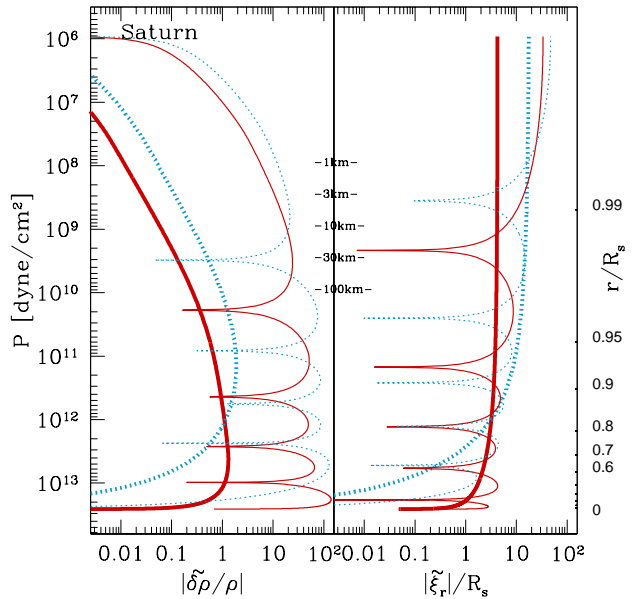}
   \caption{
Normalized eigenfunctions for a few fundamental- and p-modes
     in Saturn.  The red (or solid) lines are for the 
     $\ell=2$ modes and the blue (or dotted) for $\ell=10$;
heavy lines represent f-modes, while thin lines the
     $n=5$ radial overtone p-modes.
Mode excitations by impacts are directly
     proportional to the heights of the eigenfunctions here, with the
     explosion driving related to the fractional Lagrangian density perturbation
     (left panel), and the momentum punch related to the radial displacement (right
     panel). 
     Legends in the middle indicate the approximate stopping depths
     for impactors of various radii.  }
    \label{fig:plot-mode}
\end{figure}

\begin{figure*}
        \centering
   \includegraphics[width=0.48\textwidth,trim=10 150 40 120,clip=]{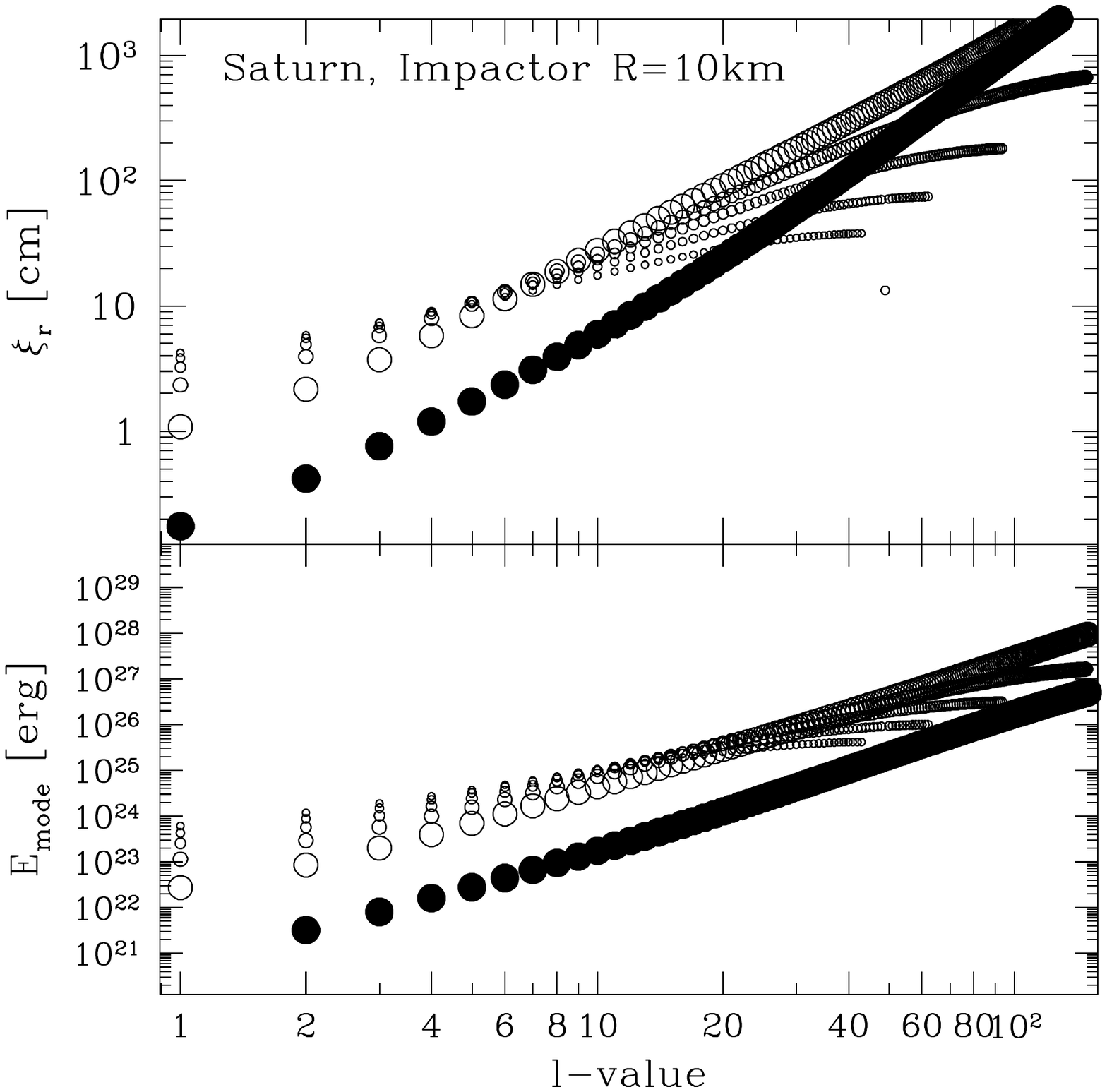}
   \includegraphics[width=0.48\textwidth,trim=10 150 40 120,clip=]{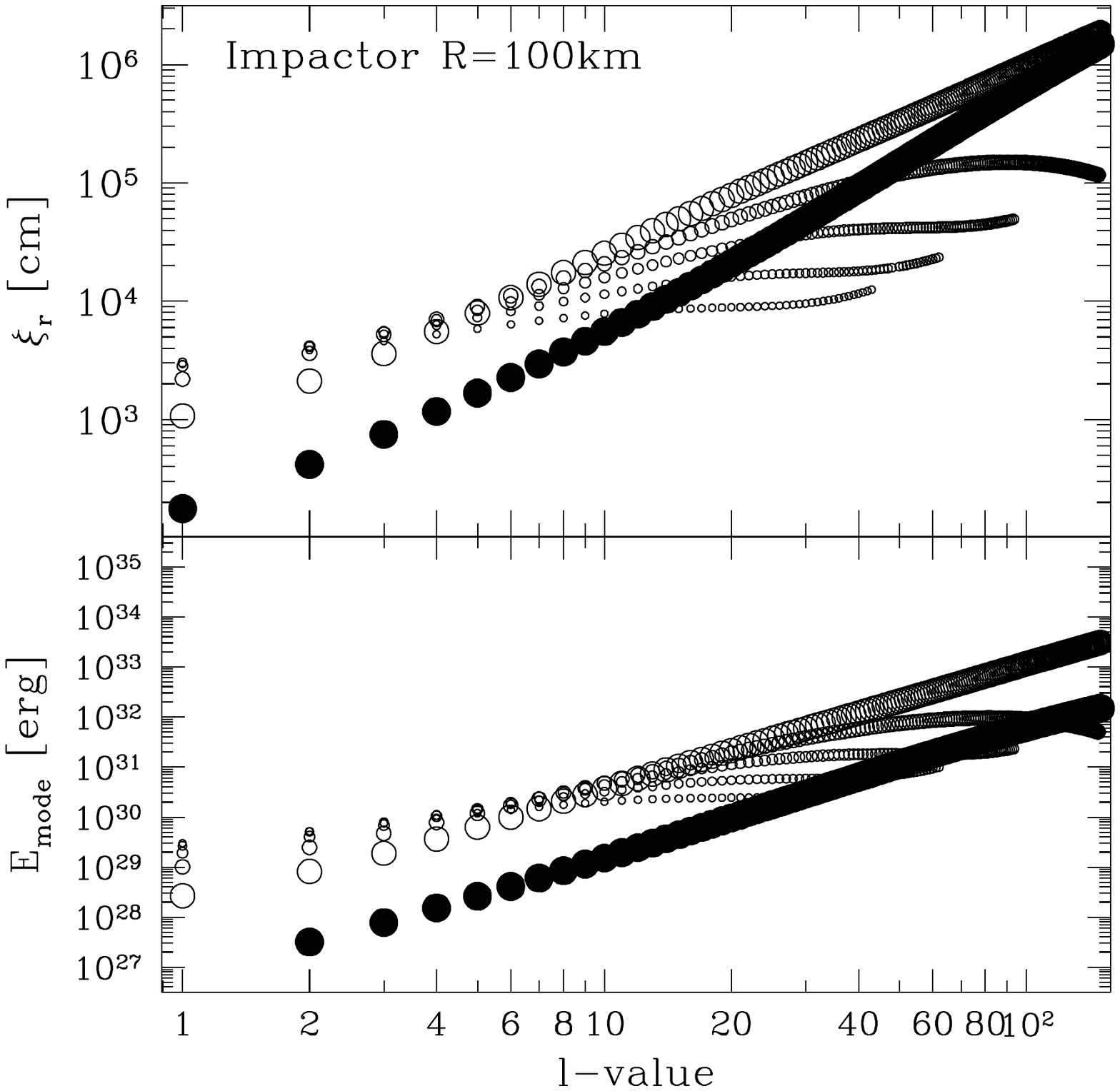}
   \caption{Mode amplitudes (in terms of surface radial displacements,
top panels) and energies
     (bottom panels) excited by an impactor of radius $R = 10\km$
     (left) and one of $R = 100\km$ (right), entering at the pole.
     F-modes are shown as solid circles and p-modes as open circles
     with smaller sizes indicating higher radial overtones. 
   }
    \label{fig:mode-excitation}
\end{figure*}

Inspired by the impact of comet Shoemaker-Levy 9 onto Jupiter,
\citet{ZahnleMaclow} simulated a 1-km body colliding with Jupiter at
the escape speed. There is a rich phenomenology (penetration,
airburst, backward outflow, fireball, chemistry...), but here we focus
only on two aspects that are of relevance to f-mode excitation.

As the body plows into the planet, it is rapidly decelerated to a halt
after encountering material of a column density roughly comparable to
itself \citep[Fig. 1 of][]{ZahnleMaclow}. There it deposits most of
its kinetic energy.
The resulting heating causes an initially supersonic expansion that
quickly transitions to subsonic in all directions, except perhaps for
the upwards direction.\footnote{Depending on the impact energy, the
  upward propagating shock wave may break through the atmosphere. This
  was observed as erupting plumes after fragments of comet
  Shoemaker-Levy 9 entered Jupiter \citep{SL9}.}
Much of the over-pressure
from the initial fireball dissipates gradually over the much longer
thermal timescale. The impactor also deposits its downward momentum at
the same location. Here, we make the optimistic assumption that the
heated bubble contains all of the impact energy as well as all of the
impact momentum.

Impacts can excite oscillations in two ways.\footnote{Normal modes can
  also be excited by (non-impacting) tidal encounters
  \citep[e.g.][]{Wu18}. We find that the magnitude is not competitive
  with direct impact, even accounting for the somewhat higher rate for
  tidal encounters.}  First, the nearly instantaneous jump in gas
pressure excites oscillations. This can be studied either by using the
over-pressure as a forcing function on an eigen-mode
\citep{Lognonne94,DB95}, or by linearly decomposing the over-pressure
signal onto a basis of eigenfunctions.
We follow the first approach here. 
Second, the momentum delivery also exerts a force on the fluid and drives
oscillations \citep{Kanamori93,Lognonne94,DB95}. We call the former
'explosion' excitation, and the latter 'momentum punch'. 
In the Appendix, we follow
\citet{Lognonne94,DB95} in deriving these two kinds of excitations,
Figure \ref{fig:mode-excitation} shows the  result produced by impactors
of two different sizes, after summing energy gains from both mechanisms.
In the following, we provide a qualitative description.

Fundamental modes are surface gravity-waves with most of their energies
stored in kinetic and gravitational potential, and only a small
fraction in compression (Fig. \ref{fig:plot-mode}).
As is shown by Fig. \ref{fig:plot-mode}, they are largely
incompressible  with
\begin{equation}
{{\delta \rho}\over{\rho}} \ll {{\xi_r}\over {R_s}}\, .
\label{eq:incomp}
\end{equation}
So they are primarily excited by the momentum punch mechanism.
P-modes, on the other hand, are compressive and can be effectively
excited also by the explosion mechanism.

Following eq. \refnew{eq:amplitude12} and eq. \refnew{eq:Escale},
the energy imparted to a mode due to the momentum punch of an impactor
entering at escape velocity is
\begin{eqnarray}
E_{\rm mode} & = & {{\sqrt{\ell(\ell+1)}}\over 2}
{{GM_s^2}\over{R_s}}\times  a_{\rm punch}^2 \nonumber \\
&\sim  &
\left[ \ell(\ell+1) {{E_{\rm impact}}\over{E_{\rm grav}}}\right] E_{\rm
  impact}\, ,
\label{eq:Escale2}
\end{eqnarray}
where $a_{\rm punch}$ is the dimensionless amplitude excited by the
punch. 
Here, $E_{\rm impact}$ is the impactor's gravitational energy and
$E_{\rm grav}$ is Saturn's binding energy, $GM_s^2/R_s$. 

Higher-$\ell$ modes are more strongly excited because they live closer
to the surface where the impact occurs.  This trend is reversed at
$\ell\geq 100$, as can be observed in the right panels of
Fig. \ref{fig:mode-excitation}. This is related to the penetration
depth of the impactor. A large impactor can deposit most of its
momentum at a large depth, below where high-$\ell$ f-modes have the
largest displacements. A similar behaviour is also observed for
acoustic modes (radial order $n > 0$). Impact excitation initially
rises with $n$ because these modes are more superficial. But by the
time $n$ is above a few, their first nodes lie shallower than the
impact depth, and their eigenfunctions oscillate over the impact
region. Impact forcing cancels to a large degree. We model this effect
using the simplistic eq.  \refnew{eq:modification}.

The scaling in eq. \refnew{eq:Escale2} also shows that larger
impactors are more efficient in exciting the oscillations. 

Mode energies scale steeply with impactor mass ($M$) and radius ($R)$
as
\begin{equation}
E_{\rm mode} \propto M^2 \propto R^6\, .
\label{eq:rscaling}
\end{equation}
Such a scaling can be observed by comparing the left and right panels
in Fig. \ref{fig:mode-excitation}.  In our calculations that include
modes up to $\ell \sim 150$, the smaller impactor converts some $3\%$
of its total energy into normal modes, while the larger one converts
$500\%$. The latter efficiency, exceeding unity, reflects the failure of
our procedure in cutting off high order, high degree modes
(eq. \ref{eq:modification}). Moreover, these modes have such high
frequencies, it is no longer possible to imagine that the impact
forcing is a delta-function in time.

We have also made a number of other simplification o produce results
in Fig. \ref{fig:mode-excitation}. We assume that the impact arrives
radially at the pole. So only the zonal modes (azimuthal quantum
number $m=0$) are excited.  Reality is more complicated: the rotation
axis can be different from the impact axis, and the impact can be
off-centre. Both give rise to a spread in mode energies between modes
with the same $n$ and $\ell$ quantum numbers, but different $m$
values. 

In summary, only a finite number of modes are effectively excited by
the impact, and larger impactors are more efficient in driving these
modes.



\begin{figure*}
        \centering
   \includegraphics[width=0.51\textwidth,trim=30 170 60 120,clip=]{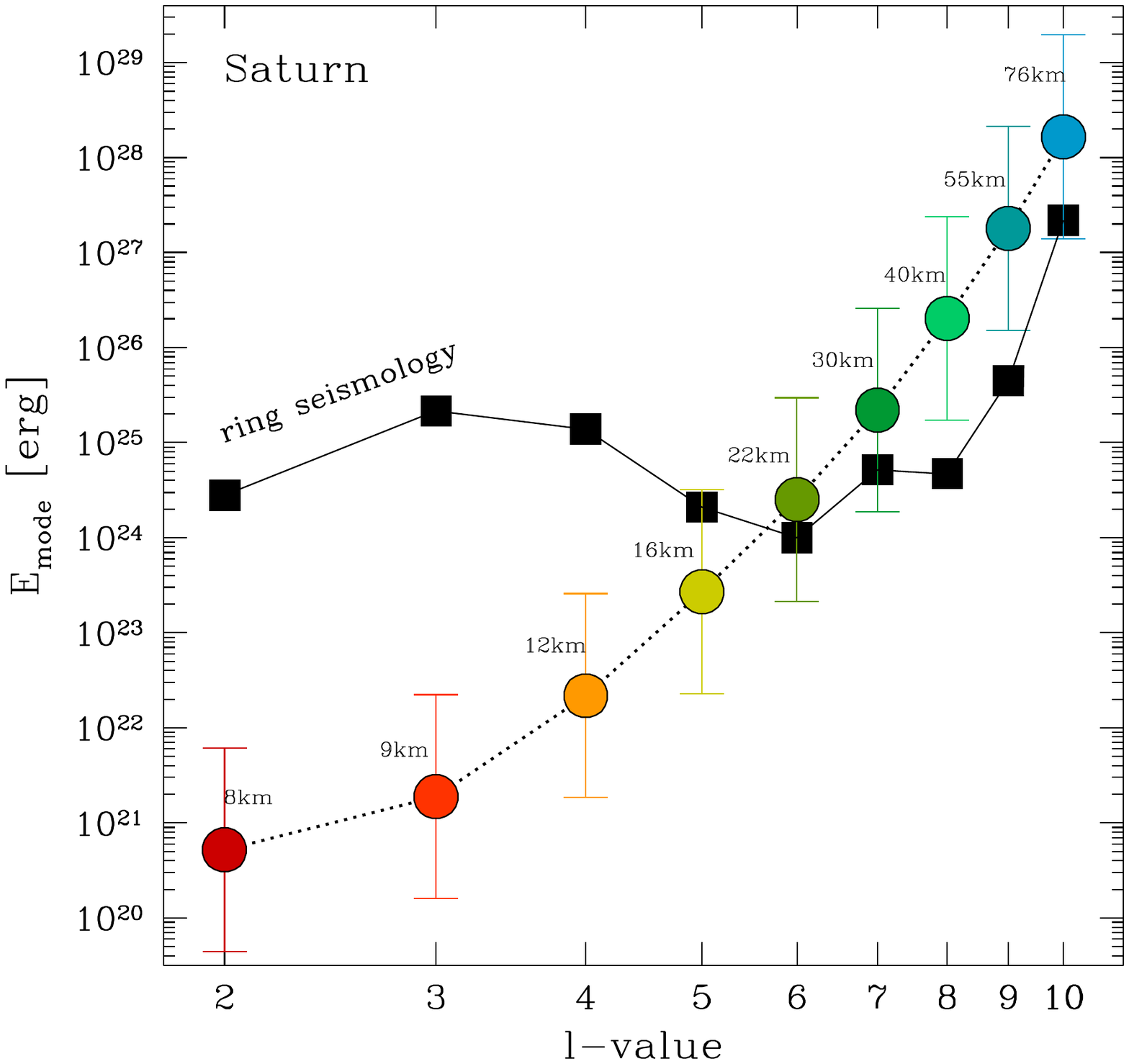}
   \includegraphics[width=0.481\textwidth,trim=60 170 60 120,clip=]{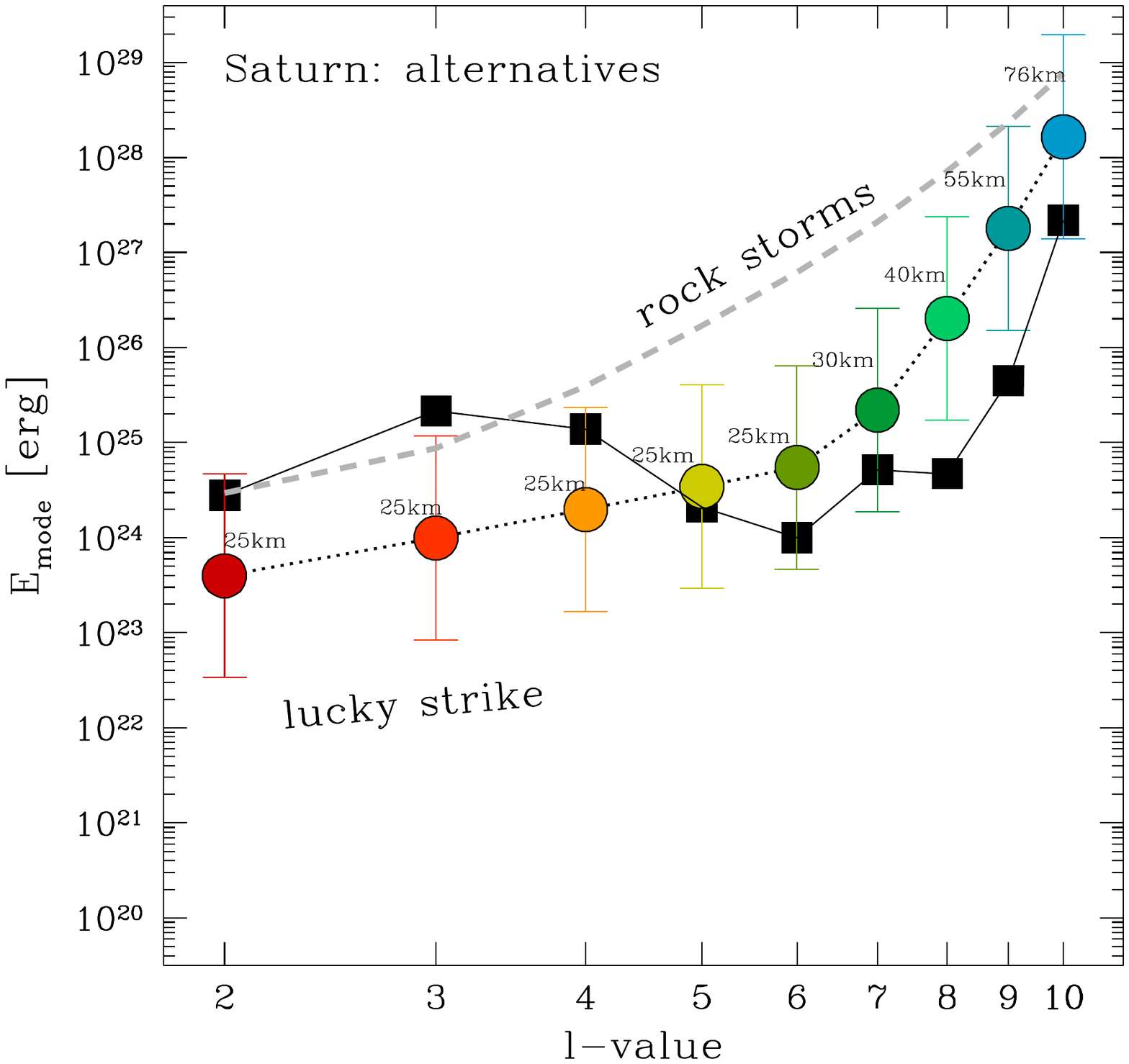}
   \caption{Energies of f-modes in Saturn.  
Mode energies derived from ring seismology are shown as solid squares.  
Color circles in the left panel present energies expected from the largest impact
     that arrive within a mode's lifetime, with small texts
     indicating the approximate radii of the responsible impactor, and
     error-bars reflecting a factor of $\sqrt{10}$ uncertainty in impact
     rates either way. Impact theory can explain high-$\ell$ modes,
     but fail by about a factor of $1000$ for the three lowest-$\ell$
     modes. To reconcile this, we present in the right panel outcomes from two
     alternative scenario, one (lucky strike, color circles) in which a
     large body ($R=25\km$) hit within the past $40,000$ yrs; the
     other ('rock storms', grey dashed curve) where deep-seated storms
     inside Saturn is responsible for exciting the f-modes.  The
     storms are assumed to carry $2\%$ of the internal flux.
   }
    \label{fig:fmode-saturn}
\end{figure*}

\subsection{Exciting  f-modes in Saturn}

We now proceed to estimate mode energies excited by past impacts on
Saturn.  

According to the observed size spectrum (eq. \ref{eq:brate}),
impactors bring in roughly equal energy flux per logarithmic size
decade.  In the mean time, larger impactors are much more efficient at
conducting their energy into the normal modes. So over a given time
window, even though the largest impactors are the rarest, they
contribute the most to exciting f-modes. 

The relevant time window for a given mode is its lifetime. As
high-$\ell$ modes are more long-lived, they have a longer memory of
past impacts, and consequently, weightier 'largest impactors'. They
should be excited to much higher energies.
Combining 
eqs. \refnew{eq:brate},
\refnew{eq:Escale2}, \& \refnew{eq:rscaling}, as well as our previous
result that $t_{\rm damp} \propto [\ell(\ell+1)]^2$, we obtain the
following scaling
\begin{equation}
E_{\rm mode} \propto \ell(\ell+1) R^6 \propto \ell(\ell+1) t_{\rm damp}^{6/2.8}
\propto \ell^{10.6}\, .
\label{eq:eexpect}
\end{equation}
Numerical results shown in Fig. \ref{fig:fmode-saturn} (left panel)
bear out this scaling, with expected mode energies varying steeply
from $10^{21} \erg$ for the $\ell=2$ mode to $10^{28} \erg$ for the
$\ell=10$ mode.  Impacts that are responsible for these modes (the
largest impacts) range from $R\sim 8\km$ for the strongly damped
$\ell=2$ mode, to $R \sim 76\km$ for the weakly damped $\ell=10$ mode.

Compared to the observed energies, our expectations are too low by
about $3$ orders of magnitude for the lowest degree modes (
$\ell=2,3,4$), but lie somewhat above (and are therefore compatible
with) those for the $\ell=5-10$ modes.



\subsection{The Low-$\ell$ problem}

Although the impact theory appears to be a partial success, its
ability to explain the observed energies in modes of the lowest
degrees is worrisome. This either implies an unknown source of
excitation for these modes, or that some aspects in the impact theory
are wrong. We briefly discuss a few possibilities here.

First, could the observed low-$\ell$ modes be the result of an
unexpectedly large invader in the recent past? To raise their energies
up by a factor of $10^3$ would require a body with size $R \sim 25\km$
(eq. \ref{eq:rscaling}), impacting Saturn within the past $40,000$
yrs. In contrast, such a body is expected to visit once every $10^{6}$
yrs (eq. \ref{eq:brate}).  So this is a small (but not vanishingly
small) probability event. And uncertainties in the impact rate (up to
a factor of $10$) could mitigate some of the discrepancy. In the
right-panel of Fig. \ref{fig:fmode-saturn}, we present the outcome of
such a 'lucky' scenario. This appears to resolve much of the tension
between theory and observation. The disadvantage of such suggestion is
that it is hard to disprove.

Another possibility is that the mode damping timescale is incorrectly
calculated. If the low-degree modes are damped much more weakly than
we have calculated, they can have longer memories of past impacts and
can be excited to larger amplitudes. It also happens that the density
waves forced by the low-degree modes are nearly nonlinear,\footnote{Visual
  inspections of the optical depth profile would in fact argue that
  the $\ell=2$ and $3$ waves are in fact nonlinear.}  as is suggested by
Fig. \ref{fig:fmode-ring}. Is it possible that nonlinear density waves
are less efficient in transporting angular momentum and energy away
from the resonant location than their linear counter-parts? Such a
possibility is not favoured by theoretical calculations
\citep{Shu,BGT}, which argue that they are as efficient as would be
expected from linear calculations.






Another suggestion is that low-$\ell$ modes may exchange energy with
modes at higher energies. As Fig. \ref{fig:mode-excitation} shows,
impacts can excite a large number of f-modes and p-modes.  The three
f-modes of concern here, with $m = \ell \in [2,3,4]$, are special in
that they are quickly damped by resonances with the rings. But they
are embedded in a heat bath where other modes oscillate with much
higher energies.  If some process, e.g., scattering by convective
eddies \citep{GoldreichMurray}, or nonlinear mode coupling
\citep{WuGoldreich}, can bring the modes toward energy equipartition,
this could help explain the energy floor at $\sim 10^{24} \erg$ for
the low-$\ell$ modes. Further investigations are needed.

Lastly, we return to the discussion on rock storms. As one such
deep-seated storm releases an energy comparable to that of a $R=15\km$
impact (\S \ref{subsec:storm}), over the lifetimes of the detected
f-modes, they may pump enough energies to explain the
observations. These are shown in the right panel of
Fig. \ref{fig:fmode-saturn} as a grey dotted curve, under the
assumption that the storms carry a few percent of Saturn's internal
luminosity (so as to reproduce the level in the $\ell=2$
mode). Whether such an assumption is physically motivated is not
known.

So to summarize, we are left at an unsatisfactory state. There are
a couple alternative scenarios but we have no means to distinguishing them.

\section{Exciting Jupiter's normal modes}
\label{sec:jupiter}

Assuming that internal dampings are weak on Jupiter, as on Saturn, the
absence of a massive ring around Jupiter brings about dramatic
consequences for the impact excitation.  Now, the bombardment history
can be remembered for of order the modes' internal damping times,
which are longer than a billion years. We expect much larger mode
amplitudes, ones that can be readily observed.\footnote{Conversely, a
  lack of large oscillations might suggest that Jupiter had a ring
  system in its not-too-distant past.}

\begin{figure}
        \centering
   \includegraphics[width=0.48\textwidth,trim=30 170 60 120,clip=]{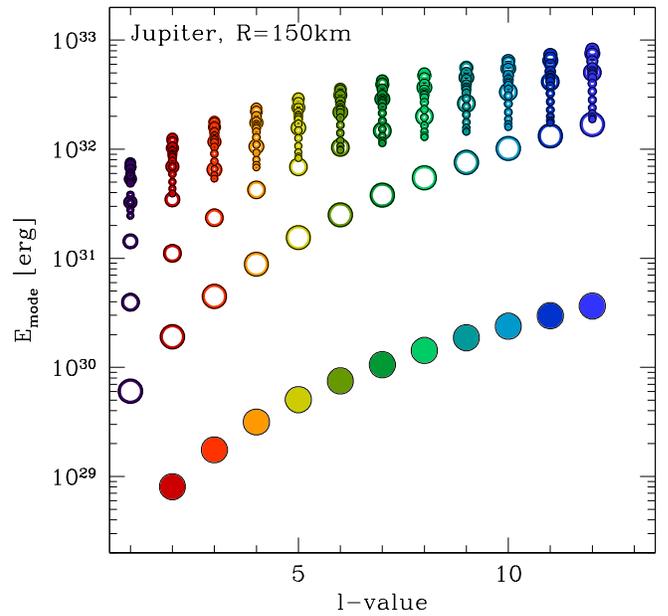}
   \caption{Modes excited on Jupiter by a single $R=150\km$
     impact. F-modes are shown as solid circles, and p-modes as open
     circles, with the circle size decreasing with radial order.  The
     $\ell=1$ f-mode does not exist in nature. P-modes of radial order
   $n \sim 3-6$ acquire the highest energies.}
    \label{fig:fmode-jupiter-energy}
\end{figure}

To illustrate, we consider a single impact of $R = 150\km$, which has
an arrival rate of $\sim 1/$Gyrs on Jupiter
\citep{SaturnRate}. Energies for some of the most excited modes are
shown in Fig. \ref{fig:fmode-jupiter-energy} -- they are much higher
than the counterparts in Saturn.  For such a large impactor, one
expects that most of the impact energy and momentum are deposited at a
pressure $p_{\rm imp} \sim 10^{10}$ cgs. This means acoustic modes
with radial order $n \sim 3-6$ achieve the highest excitation. Modes
at higher radial orders can propagate to above this depth, and their
excitation suffers cancellation.

\begin{figure*}
        \centering
   \includegraphics[width=0.48\textwidth,trim=30 170 60 120,clip=]{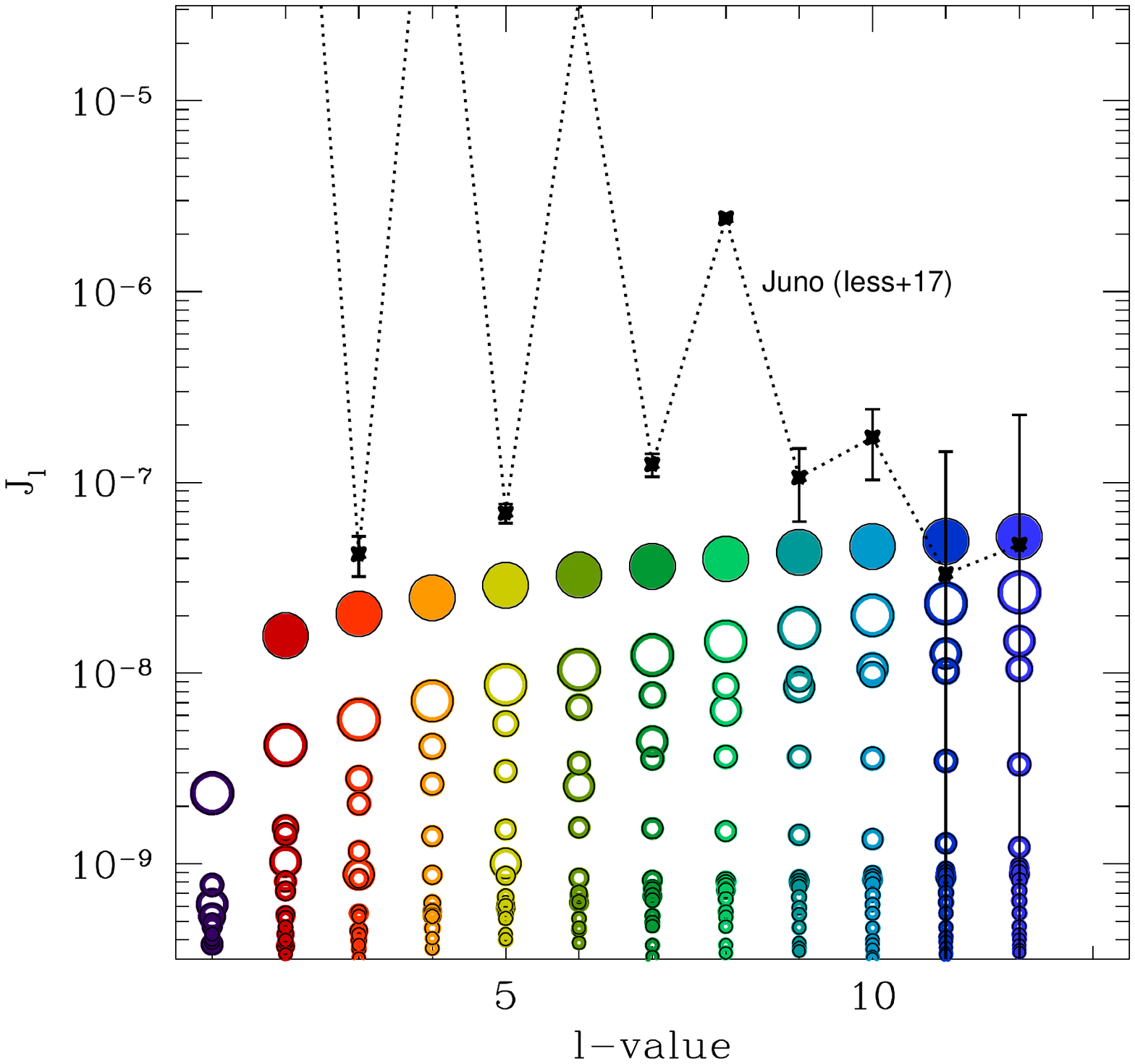}
   \includegraphics[width=0.48\textwidth,trim=30 170 60 120,clip=]{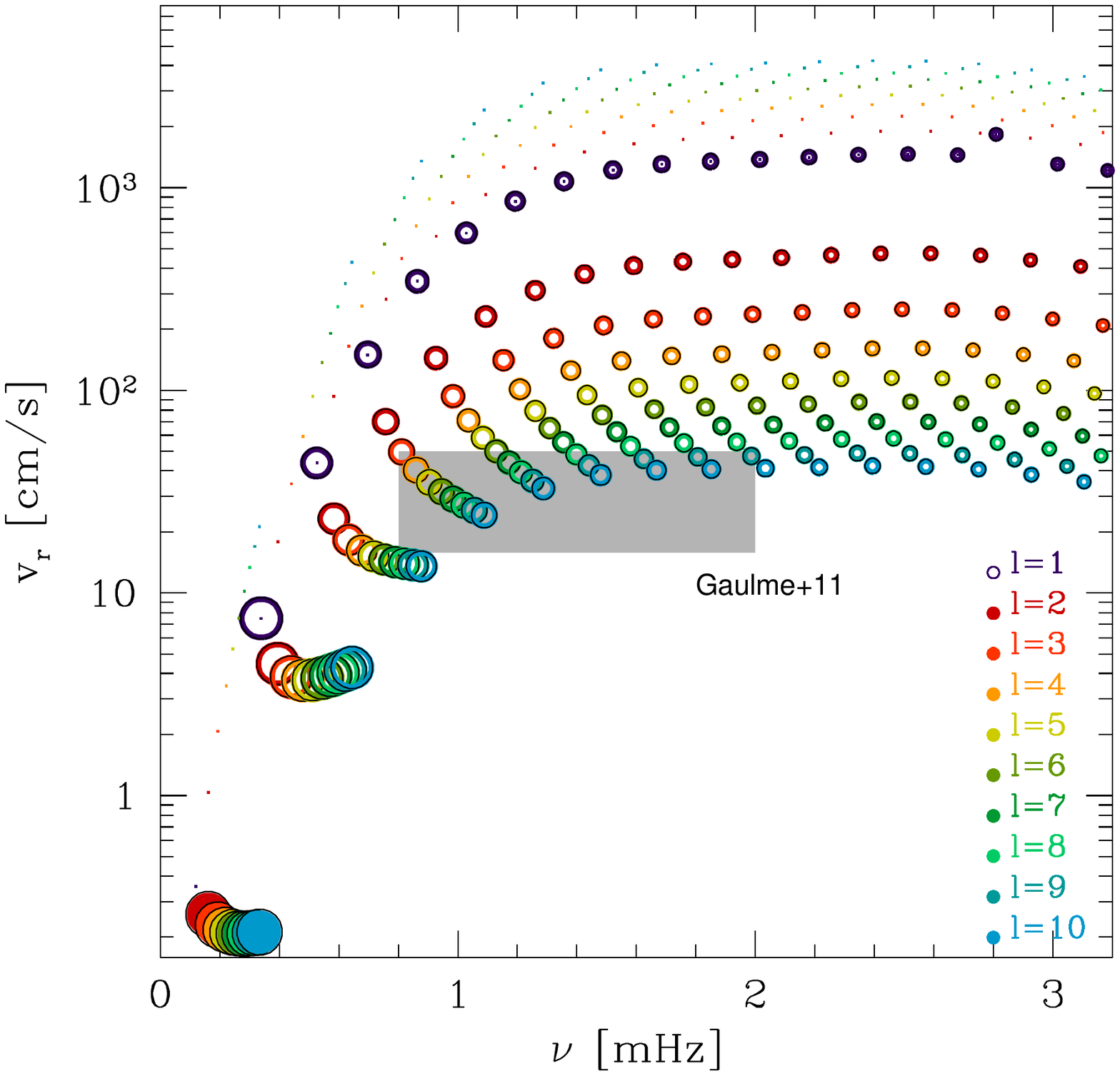}
   \caption{Observable signatures, after an impact by a
     $R=150\km$ body on Jupiter. Symbols are the same as in
     Fig. \ref{fig:fmode-jupiter-energy}. 
The left panel depicts the associated zonal ($m=0$) gravitational moments.
Black points with error bars indicate {\it Juno} measurements for the
     static gravitational moments of Jupiter after a few 
     crossings \citep{Iess17}. 
The right panel displays the expected surface radial velocities,
with small dots marking the maximum velocity amplitudes at
surface, and circles those integrated over the visible disk (see text).
The grey box marks the strength and frequency of detections
sas reported by \citet{Gaulme11}.
}
    \label{fig:fmode-jupiter}
\end{figure*}

In the following, we translate these energies into two observables:
zonal gravitational moments, $J_\ell$ (eq. \ref{eq:JL}), potentially
measurable by the {\it Juno} spacecraft currently orbiting Jupiter;
and surface radial velocities $v_r$, detectable by ground-based
spectroscopy \citep{Mosser00,Gaulme11}. The following results for
$J_l$ and $v_r$ both scale with the impactor size as $R^3$.

The expected gravitational moments are shown in the left panel of
Fig. \ref{fig:fmode-jupiter}. F-modes, being more deep-seated,
dominate the signal, despite their lower energies. These are
contrasted against  current measurements.  The static
(zero-frequency) zonal moments of Jupiter have been measured down to a
precision of order $10^{-8}$, after just a few swing-bys of the {\it
  Juno} spacecraft \citep{Bolton17,Iess17}.  While the even order
zonal moments ($J_2, J_4...$) are large and are dominated by
rotational flattening of Jupiter, the odd moments ($J_3, J_5...$),
likely caused by the surface zonal winds, have magnitudes of order
$10^{-7}$.  These are comparable to our estimates for the
impact-excited f-modes. So the prospect is bright if a similar precision
can be reached for the oscillating moments by analyzing the Doppler
residual in {\it Juno's} orbit\citep{Durante}.
Different from the zonal winds, f-mode signatures are both
time-dependent and may include tesseral moments ($m\neq 0$).

P-modes, instead of f-modes, dominate the radial velocity signal. We
find that the maximum signal should lie around $1-3$ mHz, and
feature p-modes of radial order $5-20$ (right-panel of
Fig. \ref{fig:fmode-jupiter}). The maximum amplitudes on the surface
of Jupiter can reach of order a few tens of metres per second, with
higher-$\ell$ modes having larger amplitudes. However, when one
observes the un-resolved Jupiter from the ground, the radial velocity
signals are those projected along the line-of-sight, and averaged over
the visible disk. For our order-of-magnitude exercise, we approximate
these by dividing the
maximum velocities by a factor of
$\ell^2$, to account, crudely, for the cancellation between
different light and dark patches. 
This exercise diminishes the signals for high-$\ell$ modes, and we
predict that the radial velocities are now dominated by $\ell=1$ modes,
peaking at $v_r \sim 10\m/\s$.

Using a dedicated radial velocity instrument and observing the solar
Mg line at 517 nm reflected off Jupiter's clouds, \citet{Gaulme11}
reported excess power on Jupiter, at frequencies between $0.8$ and
$2.1$ mHz.  and a possible secondary excess power between $2.4$ and
$3.4$ mHz, just below the acoustic cut-off frequency at $3.5$ mHz.
Encouragingly, these signals are modulated by a comb-like structure,
resembling those of p-modes with $n=4-11$, and $\ell \sim 1$.
The frequency and velocity amplitude they reported are crudely shown
as a grey-box in Fig. \ref{fig:fmode-jupiter}, where the amplitudes
are obtained using a $3\mu Hz$ window smoothing. 
Fitting the data
using a single sine-wave (a delta-function in frequency), they set a
limit of $50\cm/\s$ for a single mode. 
%

So while the frequency and mode identity reported by \citet{Gaulme11}
appear consistent with what we expect from impact excitations, their
maximum velocity is a factor of $\sim 20$ too low. Part of this may be
related to our simplified, order-of-magnitude-level calculations, but
a bigger part may be related to a problem we have not yet tackled --
the internal damping of p-modes. These modes are compressible and
propagate very close to the surface. They also have frequencies close
to the acoustic cut-off. 
\citet{Mosser95} have studied this problem and argued that p-modes
with frequencies between $1-2$ mHz are strongly damped by wave leakage
with $Q \leq 10^7$. If this is indeed true, they can dramatically reduce our
predicted amplitudes for these modes. 

In summary, Jovian f-modes should be excited to levels likely
detectable by the {\it Juno} mission, while impact-driven low-order
p-modes may explain the radial velocity signals reported by
\citet{Gaulme11}. More investigations are needed. 

It is interesting to note that Saturn may also have detectable radial
velocity signals -- only a few f-modes have resonances in the rings
and are thus strongly damped. Other modes should have much higher
amplitudes.

\section{Conclusions and Predictions}
\label{sec:conclusion}

F-modes are excited in Saturn. They force density waves in Saturn's
rings that become visible in the {\it Cassini} stellar occultation
data. We have obtained a number of theoretical results regarding the
driving and damping of these modes:
\begin{enumerate}

\item the very density waves that reveal these oscillations remove
  energy from them on a timescale that ranges from $10^4$
  to $10^7$ yrs;

\item in comparison, any other energy sink inside Saturn is negligible
  for these f-modes, with damping timescales that run upward of
  billions of years;

\item unlike convection in the Sun, turbulent convection in Saturn is
  too feeble to excite these modes to the observed level; so are
  surface storms that are driven by latent heat release of water
  (``water storms''); deeper storms that are possibly driven by the
  latent heat of refractive elements (``rock storms'') may be potent
  in driving these modes, but their existence and frequency remain
  unknown;

\item another source of stochastic excitation is impacts. The long
  damping times allow the modes to remember large impacts from eons
  ago.  High-$\ell$ modes are excited to higher energies as they have
  longer memories. This theory adequately explains the observed
  amplitudes in the $\ell=5-10$ f-modes, but falls short for the
  lower-$\ell$ modes;

\item the low-$\ell$ modes may be excited by a fortuitously large
  recent impact; or a different mechanism (e.g., 'rock storms'); or
  energy exchange with other modes; 

\item transcribing these processes to the case of Jupiter, which does
  not have a massive ring, suggests that Jupiter could still be
  vibrating from very large impacts that arrived billions of years
  ago; this opens the possibility of direct detection by the {\it
    Juno} spacecraft, and of explaining literature claims of
  ground-based detections.

\end{enumerate}

There are many caveats and puzzles that we fail to resolve in this
work. We list a few of them here, together with some predictions that
may be tested in the near future:

\begin{itemize}

\item it remains a coincidence that f-modes in Saturn acquire just
  enough energy to excite visible features in Saturn's C-rings. This
  would not have been possible if, e.g., the C-rings were a few times
  denser, or the impact rates were a few times lower.

\item inferred surface displacements for all f-modes cluster around
  $1\m$. Is this a coincidence?

\item density waves driven by low-$\ell$ modes are likely
  nonlinear. Could this influence our estimates for their lifetimes?

\item impacts excite f-modes over a broad range of $\ell$ and $m$
  values. They also excite acoustic modes. Only some of the f-modes,
  ones that find resonances within Saturn's rings, are easily visible
  and are drained rapidly. The rest should have larger amplitudes and
  could be searched for either in the {\it Cassini} orbital data, or
  using radial velocity techniques from the ground.

\item The acoustic modes will dominate the radial velocity signals.
  Our crude predictions for their amplitudes, in Saturn and in
  Jupiter, are based on the assumptions that they have lifetimes
  comparable to that of f-modes. This deserves further scrutiny.
  Detecting these modes is valuable because, compared to f-modes, they
  are more sensitive to the interior structure of the giant planets.

\end{itemize}

The Earth and Moon record past bombardments, in the form of craters on
their surfaces. Giant planets can also remember their history, in the
form of long-lived oscillations.


\bigskip

{\it acknowledgement} We wish to acknowledge discussions with Phil
Nicholson, Matt Hedman, Peter Goldreich and Yuri Levin. WYQ thanks
NSERC for research funding, YL acknowledges NSF grant AST-1352369 and
NASA grant NNX14AD21G.


\begin{appendix}






\section{Impact Excitation}

To calculate the excitation of internal modes by impact, we adopt the
derivations in \citet{Lognonne94,DB95}, recapped here briefly.

Starting from the fluid equation of motion,
\begin{equation}
\rho {{d {\boldv}}\over{dt}} = - {\boldnabla} P + \rho {\boldnabla} \Phi
+ {\boldF}(\boldr,t)\, 
\label{eq:eom}
\end{equation}
where the last term represents the forcing related to the impactor, we
decompose the forced response in terms of the free oscillation modes,
${\boldxi}_k ({\boldr})$,
\begin{equation}
{\boldxi}({\boldr},t) = \sum_k a_k(t) {\tilde \boldxi}_k({\boldr})\, ,
\label{eq:decompose}
\end{equation}
where the eigenfunctions are normalized as in eq. \refnew{eq:mynorm}.
In particular, the radial displacement for mode $k$ has the form 
\begin{equation}
{\tilde \xi}_r(\boldr) = {\tilde \xi}_r (r) Y_{\ell m}(\theta,\phi)\, ,
\label{eq:xirylm}
\end{equation}
and the spherical harmonics function is normalized as $\int d\Omega
Y_{\ell m} Y_{\ell m}^* = 1$.

We can recast the equation of motion into the following simpler
form
\begin{equation}
{\ddot{a}}_k + \omega_k^2 a(t) = F_k (t)\, ,
\label{eq:at}
\end{equation}
where 
\begin{equation}
F_k (t) = {{\int dV \, {\boldF}({\boldr},t) \cdot {\tilde \boldxi}_k ({\bf
      r})}\over{\int dV \rho(r) {\tilde \boldxi}_k \cdot {\tilde \boldxi}_k}}\, .
\label{eq:Fk}
\end{equation}
The response of the mode to a slowly varying forcing ($F_k(t)$) is
facilitated by adopting a new variable that is slowly varying in time,
$b_k(t) = a_k(t) e^{-i \omega_k t}$. This allows us to ignore ${\ddot
  b}_k$ and obtain
\begin{equation}
b_k (t)  \approx -{{i}\over{\omega_k}} \int_{-\infty}^{\infty} F_k(t) e^{-i
  \omega_k t} dt\, .
\label{eq:bk}
\end{equation}

We now consider two possible forcings exerted by an impact, the
over-pressure region after the impactor's energy is converted into
heat ('explosion'), and the momentum the impactor deposited at the
envelope ('punch'). We make a number of simplifications to describe the
temporal and spatial behaviours of the forcing.  As the impactor
travels fast and is braked after encountering a comparable amount of
mass, the event only lasts $\Delta \tau \sim l/v \sim$ a few seconds,
where $l$ is the penetration depth. So the first one can be thought of
as a rapid deposition followed by a slow dissipation, as the local
thermal time is long,
or, roughly, a Heaviside step function ${\cal H}(t)$.  The momentum
punch, on the other hand, can be thought of as a top-hat function with
duration $\Delta \tau$.  Spatially, we assume that the explosion can
be described as an over-pressure ($\delta P_{\rm expl}$) region at
location ${\boldr}_0$ with a radius of $\Delta r$, and the total momentum
($\delta {\bf p}_{\rm punch}$) is deposited in the same region, both
again approximated by the Heaviside function,
\begin{eqnarray}
& {\boldF}_{\rm expl} (\boldr,t) & = - {\boldnabla} \left\{ {\delta P_{\rm
    expl}}\times  
{\cal H}\left[ {{|{\boldr} - {\boldr}_0|}\over{\Delta r}}- 1\right]\right\}
\times {\cal H}(t)\, \\
&{\boldF}_{\rm punch}(\boldr,t) &= {{\delta \boldp_{\rm punch}}\over{\Delta V \Delta \tau}}
\times 
{\cal H}\left[ {{|{\boldr} - {\boldr}_0|}\over{\Delta r}}- 1\right]
\times \left[{\cal H}(t+\Delta \tau)  - {\cal H}(t)\right]\, .
\label{eq:forcing}
\end{eqnarray}
where $\Delta V = 4\pi/3 \Delta r^3$ and $\Delta r$ is likely of order the local
pressure scale height. The Heaviside function has the property of
$\int_{-\infty}^{\infty} {\cal H}(t) e^{-i\omega_k t}dt = {1\over{i
    \omega_k}}$.  In the following, we assume that the impact region
is small and shallow, $\Delta r \ll R_s$, $r_0 \approx R_s$, and that the
impact is quick, $\Delta \tau \ll 2 \pi/\omega_k$. Moreover, we assume
that the eigenfunction hardly varies over the impact bubble. We adopt
a perfect efficiency, $\delta P_{\rm expl} \approx E_{\rm
  impact}/({4\pi/3 \Delta r^3})$, and the momentum is related to the energy by
the surface escape velocity, $\delta \boldp_{\rm punch} \approx E_{\rm
  impact}/v_{\rm esc} \boldn$.  Substituting the above expressions
into eqs. \refnew{eq:Fk} \& \refnew{eq:bk}, we obtain the final
amplitudes as
\begin{eqnarray}
&|a_k|_{\rm expl} & = {{E_{\rm impact}}\over{\omega_k^2 M_s R_s^2}}\times
{\boldnabla \cdot {\tilde \boldxi}}_k|_{{\boldr}={\boldr}_0} \,
,\nonumber \\
&|a_k|_{\rm punch} & =   {{|\delta \boldp_{\rm punch}|}\over{ \omega_k  M_s R_s^2}}
\times {\rm sinc}({{\Delta \tau \omega_k}\over 2})\times \left({\tilde
    \boldxi}_k\cdot
  {\bf n}\right)|_{\boldr = {\boldr}_0}\approx 
{{E_{\rm impact}}\over{v_{\rm esc} \omega_k M_s R_s^2}}\times
\left({\tilde \boldxi}_k\cdot
  {\bf n}\right)|_{\boldr = {\boldr}_0}\, .
\label{eq:amplitude12}
\end{eqnarray}
For f-modes, the tangential and radial displacements are comparable in
magnitude, so we can consider only the radial momentum, ${\bf n} =
{\hat \boldr}$.
Furthermore, without loss of generality, we assume that the impactor
strikes radially on the pole. In this frame, only the $m=0$ modes are
excited and we can insert $Y_{\ell 0}(\theta=0,\phi)
=\sqrt{(2\ell+1)/4\pi}$ into eq. \refnew{eq:xirylm}.

For modes with high spherical degree ($\ell$) and radial order ($n$),
the assumption that the eigenfunction is nearly constant over the
impact bubble is no longer satisfied. We may continue to employ
eqs. \refnew{eq:amplitude12}, but with the understanding that the
eigenfunction appearing on the right-hand side should be that averaged
over the bubble depth, e.g.,
\begin{equation} 
\boldnabla \cdot {\tilde \boldxi}_k \rightarrow 
 {1\over {2\Delta r}} \int_{r_0 - \Delta r}^{r_0+\Delta r}\, dr\, \boldnabla \cdot {\tilde \boldxi}_k 
\, ,\hskip1.0in 
{\tilde \boldxi}_k \rightarrow 
 {1\over {2\Delta r}} \int_{r_0 - \Delta r}^{r_0+\Delta r}\, dr\,  {\tilde \boldxi}_k 
\label{eq:modification}
\end{equation}
In practice, we take $\Delta r$ to be the local scale height. For
modes with very high spherical degree, the horizontal wavelength may
be smaller than $\Delta r$. For these, we have to perform 3-D
averaging over the bubble, further weakening the excitation.

In simulations by \citet{ZahnleMaclow}, plumes were produced by the
upward propagating shock of the impactor explosion, and they may carry
up to $35\%$ of the impact energy. Could the momentum of these plumes,
as they fall back to the planet, excite the modes further than
estimated above? The typical upward velocity is a fraction of the
surface escape velocity, $v \sim v_{\rm esc}/10$,  and so we find that the
momentum from the plume fall-back will exceed the momentum of the
original impact, by a factor of $3$.
The duration of the fall-back is short compared to f-mode periods, and
the nearly incompressible f-modes feel the momentum punch more
effectively (than over-pressure). This could boost the energy of the
f-modes by a factor of $10$, but unlikely more. 

\end{appendix}

\end{document}